\documentclass[twocolumn,floatfix]{aastex631}

\newcommand{\ra}[1]{\renewcommand{\arraystretch}{#1}}

\usepackage{amsfonts}
\usepackage{booktabs}
\usepackage{savesym}
\savesymbol{tablenum}
\savesymbol{splitbox}
\usepackage{siunitx}
\restoresymbol{SIX}{tablenum}

\usepackage{float}
\usepackage{adjustbox}
\restoresymbol{SIX}{splitbox}
\usepackage{rotating}
\usepackage{gensymb}
\usepackage{array,multirow,graphicx}

\shorttitle{$L-\sigma$ Relation in Sp1149}
\shortauthors{Williams et al.}

\graphicspath{{./}{ }}

\begin{document}

\title{Sp1149 I: Constraints on the Balmer $L-\sigma$ Relation for \ion{H}{2} Regions in a Spiral Galaxy at Redshift $z=1.49$ Strongly Lensed by the MACS J1149 Cluster}

\date{September 2023}

\author[0000-0002-1681-0767]{Hayley Williams}
\affiliation{School of Physics and Astronomy, University of Minnesota 
116 Church Street SE, Minneapolis, MN 55455 USA 
}   
\author[0000-0003-3142-997X]{Patrick Kelly}
\affiliation{School of Physics and Astronomy, University of Minnesota 
116 Church Street SE, Minneapolis, MN 55455 USA 
}

\author[0000-0003-1060-0723]{Wenlei Chen}
\affiliation{School of Physics and Astronomy, University of Minnesota 
116 Church Street SE, Minneapolis, MN 55455 USA 
}

\author[0000-0001-9065-3926]{Jose Maria Diego}
\affiliation{IFCA, Instituto de F\'{i}ısica de Cantabria (UC-CSIC), Av. de Los Castros s/n, 39005 Santander, Spain}

\author[0000-0003-3484-399X]{Masamune Oguri}
\affiliation{Center for Frontier Science, Chiba University, 1-33 Yayoi-cho, Inage-ku, Chiba 263-8522, Japan}
\affiliation{Department of Physics, Graduate School of Science, Chiba University, 1-33 Yayoi-Cho, Inage-Ku, Chiba 263-8522, Japan}

\author[0000-0003-3460-0103]{Alexei V. Filippenko}
\affiliation{Department of Astronomy, University of California, Berkeley, CA 94720-3411, USA}

\begin{abstract}
The luminosities and velocity dispersions of the extinction-corrected Balmer emission lines of giant \ion{H}{2} regions in nearby galaxies exhibit a tight correlation ($\sim0.35$~dex scatter). 
There are few constraints, however, on whether giant \ion{H}{2} regions at significant lookback times follow an $L-\sigma$ relation, given the angular resolution and sensitivity required to study them individually.  
We measure the luminosities and velocity dispersions of H$\alpha$ and H$\beta$ emission from 11 \ion{H}{2} regions in Sp1149, a spiral galaxy at redshift $z=1.49$ multiply imaged by the MACS\,J1149 galaxy cluster. Sp1149 is also the host galaxy of the first-known strongly lensed supernova with resolved images, SN Refsdal. 
We employ archival Keck-I OSIRIS observations, and newly acquired Keck-I MOSFIRE and Large Binocular Telescope LUCI long-slit spectra of Sp1149. 
When we use the {\tt GLAFIC} simply parameterized lens model, we find that the H$\alpha$ luminosities of the \ion{H}{2} regions at $z=1.49$ are a factor of $6.4^{+2.9}_{-2.0}$ brighter than predicted by the low-redshift $L-\sigma$ relation we measure from Very Large Telescope MUSE spectroscopy. If the lens model is accurate, then the \ion{H}{2} regions in Sp1149 differ from their low-redshift counterparts. 
We identify an \ion{H}{2} region in Sp1149 that is dramatically brighter (by $2.03\pm0.44$ dex) than our low-redshift $L-\sigma$ relation predicts given its low velocity dispersion. Finally, the \ion{H}{2} regions in Sp1149 are consistent, perhaps surprisingly, with the $z\approx 0$ star-forming locus on the Baldwin-Phillips-Terlevich diagram.
\end{abstract}

\section{Introduction}

\begin{figure*}[ht]
\centering
\includegraphics[width=1\linewidth]{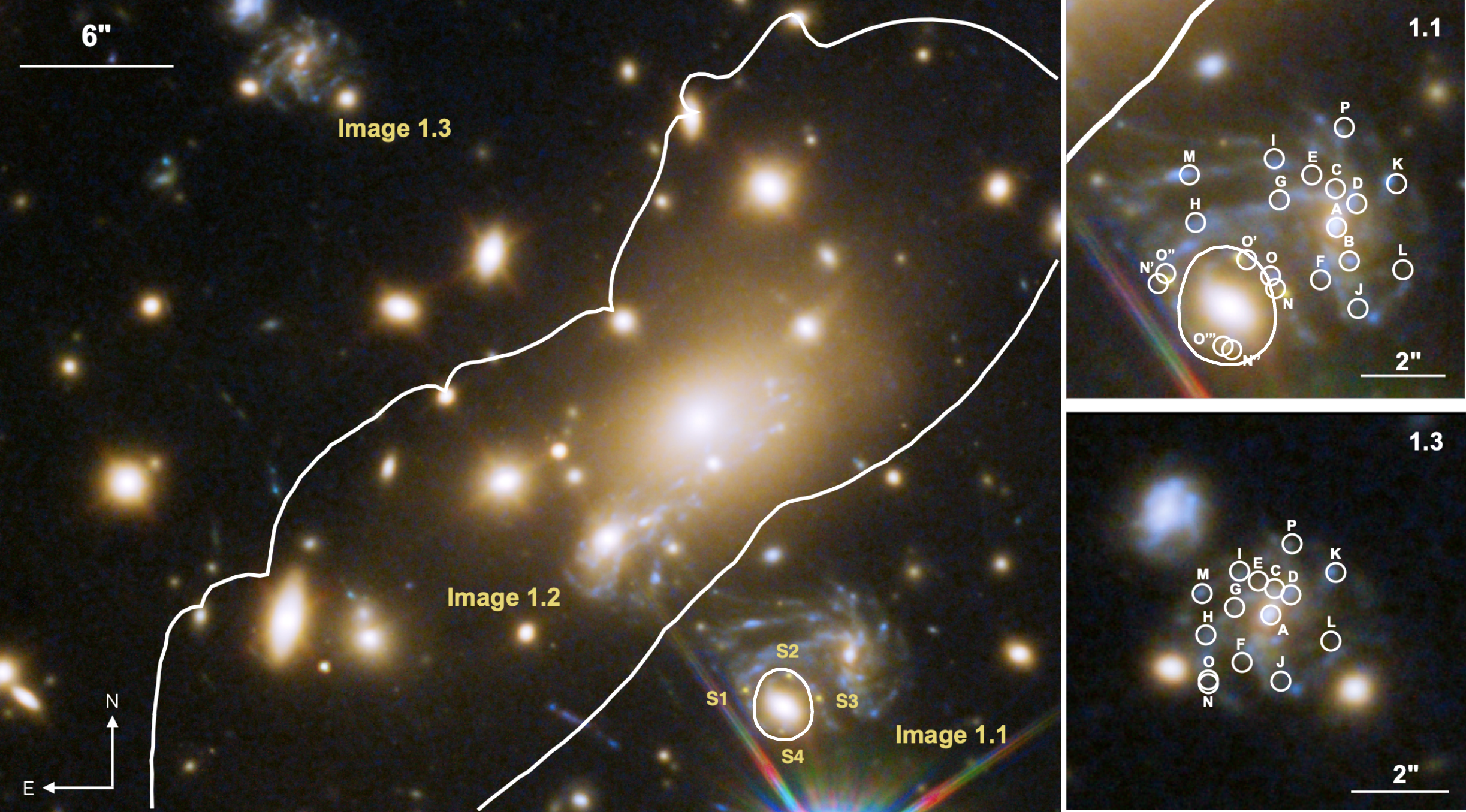}
\caption{{\it HST} imaging of the MACS\,J1149 cluster, with multiple images of the lensed spiral galaxy Sp1149. The left panel shows the critical curve of the cluster as a white line, and the critical curve of the cluster-member galaxy around which the images of SN Refsdal form an Einstein cross, taken from the GLAFIC model. The four appearances of SN Refsdal in Image 1.1 are labeled as S1--S4. The right panels show close-up views of Images 1.1 and 1.3, with the \ion{H}{2} regions in each image circled and labeled.}
\label{fig:macsj1149}
\end{figure*}


It is well established that the cosmic star-formation rate (SFR) volume density and supermassive black hole (SMBH) accretion rate peaked near redshift $z\approx 1.5$--2, an epoch nicknamed ``cosmic noon" \citep[e.g.,][]{Dickinson_2003,Hopkins_2007,Delvecchio_2014,Madau_2014}. Star-forming galaxies (SFGs) during this epoch were creating stars and accreting onto their central SMBHs about 10 times more rapidly, on average, than galaxies today \citep[e.g.,][]{Oliver_2010,Whitaker_2014,Forster_Schreiber_2020}. Observations of galaxies at cosmic noon have been crucial to our understanding of the complex feedback mechanisms, such as stellar winds and merger events, that drive the cosmic evolution of galaxies \citep[e.g.,][]{Somerville_2015}. 

Modern imaging and spectroscopy have enabled measurements of the global properties of galaxies near the peak of star formation at cosmic noon. Scaling relations of these properties, including a linear relationship between SFR and stellar mass and additional relations involving galaxy size, metallicity, and gas content, have been observed out to $z\approx 2$--3 \cite[e.g.,][]{Speagle_2014,Maiolino_2019,Tacconi_2020}. High-resolution imaging of SFGs has also revealed that galaxies at significant lookback times appear increasingly more irregular and ``clumpier"  \citep{Brinchmann_1998}, which recent observations by the {\it James Webb Space Telescope (JWST)} have confirmed \citep{Treu_2023}.

While the global properties of galaxy populations at cosmic noon are increasingly well characterized by observations, the study of individual star-forming and \ion{H}{2} regions in galaxies at $z\gtrsim 1$ has been limited by their small angular size and faint intrinsic brightness. Even with the assistance of adaptive optics (AO) on the largest telescopes, spatial resolution is limited to $\sim 1.5$~kpc at $z\approx 2$, making it impossible to resolve and study individual \ion{H}{2} regions. Magnification by gravitational lensing can help solve this problem, as the stretching of images of background galaxies provides enhanced spatial resolution down to $\sim100$~pc scales \citep[e.g.,][]{Wuyts_2014,Wang_2017,Curti_2020}. Observations of lensed galaxies can allow us to compare directly the properties of \ion{H}{2} regions in galaxies at $z\approx 1$--2 to those of \ion{H}{2} regions in the local Universe, enhancing our understanding of the processes which govern star formation at cosmic noon. Previous studies of star-forming clumps in resolved magnified galaxies have found a positive redshift evolution in surface brightness, SFR surface density, and luminosity-function cutoffs out to at least $z\approx 3$ \citep[e.g.,][]{Swinbank_2009,Jones_2010,Livermore_2012,Livermore_2015,Claeyssens_2023}.
However, given the short lifetimes of the luminous O-type stars, the positions of \ion{H}{2} regions do not all correspond to clumps identified from broadband imaging.

There is a well-established empirical relationship between the H$\beta$ luminosity and velocity dispersion of giant \ion{H}{2} regions at low redshift, the $L-\sigma$ relation, which extends to \ion{H}{2} galaxies \citep[][]{Terlevich_1981,Terlevich_Melnick_1988,Chavez_2014,Fernandez_Arenas_2018}. The low-redshift $L-\sigma$ relation applied to \ion{H}{2} galaxies is consistent with luminosity distances expected for standard cosmological parameters to $z\approx 4$ \citep{Melnick_2000,Siegel_2005,Plionis_2011,Chavez_2012,Chavez_2016,tsiapibasilakosplionis21} . However, observational challenges have made it difficult to measure the Balmer luminosities and velocity dispersions of individual giant \ion{H}{2} regions at $z\gtrsim 1$. The $L-\sigma$ relation in nearby galaxies is thought to reflect the complex dynamics and gravitational instabilities in star-forming regions, so evaluating the redshift evolution of the $L-\sigma$ relation for individual \ion{H}{2} regions would improve our understanding of the physical processes that govern star formation at cosmic noon. \cite{Livermore_2015} used integral field unit (IFU) spectroscopy to  measure the H$\alpha$ luminosities and intrinsic velocity dispersions of 50 total giant \ion{H}{2} regions from a sample of 17 magnified galaxies at  $z\approx 1$--3. They did not find a correlation in their sample between the Balmer velocity dispersion and luminosity of the lensed \ion{H}{2} regions. The authors used a global value for the extinction due to dust in each galaxy, rather than measuring the extinction for each individual \ion{H}{2} region in their sample, which might contribute to the large scatter in their measurements.

Magnified galaxies that are stretched by comparable factors in orthogonal spatial directions are the best candidates for spatially resolved studies, since the images are relatively undistorted and \ion{H}{2} regions can be accurately identified and extracted with minimal contamination from background sources.
One such galaxy is Sp1149, the host galaxy of the first known multiply-imaged supernova (SN), dubbed ``SN Refsdal" \citep{Kelly_2015} and a highly magnified ($\mu\approx 600$) blue supergiant star known as ``Icarus" \citep{Kelly_2018}. First identified in {\it Hubble Space Telescope (HST)} imaging of the MACS\,J1149.5+2223 cluster ($z=0.544$, hereinafter referred to as MACS\,J1149), Sp1149 is a multiply-imaged and highly-magnified face-on spiral galaxy at $z=1.49$ \citep{Smith_2009,Teodoro_2018}. The three full images of Sp1149 are some of the largest images ever observed of a spiral at $z>1$, and the relative lack of distortion due to lensing in the most highly-magnified image makes it ideal for spatially resolved studies (see Figure~\ref{fig:macsj1149}). IFU spectroscopy of Sp1149 was obtained in 2011 \citep{Yuan_2011} with the OH-Suppressing Infra-Red Spectrograph (OSIRIS) on the Keck~I 10~m telescope \citep{Larkin_2006}. Additional OSIRIS spectroscopy was obtained in 2015 \citep{Yuan_2015}, and the H$\alpha$ map from these combined observations revealed more than ten resolved and highly magnified \ion{H}{2} regions.

Here, we present constraints on the Balmer $L-\sigma$ relation for individual giant \ion{H}{2} regions in Sp1149 at $z=1.49$. Using the archival OSIRIS IFU observations of Sp1149 and new multislit observations from the Multi-Object Spectrometer For Infra-Red Exploration (MOSFIRE) on Keck~I \citep{McLean}, we measure the intrinsic H$\alpha$ luminosities and velocity dispersions of 11 resolved \ion{H}{2} regions in Sp1149. We determine a new low-redshift calibration for the $L-\sigma$ relation using the same physical aperture sizes, spatial resolution, and spectral resolution as used in the Sp1149 analysis, enabling an accurate comparison between the \ion{H}{2} regions at $z=1.49$ and those at $z\approx 0$. 

Throughout this work, we use a simply-parameterized lens model of the MACS\,J1149 cluster created by the GLAFIC team \citep{Oguri_2010,Kawamata_2016} to compute magnification due to gravitational lensing. The model uses the positions of 108 multiple-images of 36 different systems as constraints, and the presence of bright knots in the images of Sp1149, which straddle the critical curve of MACS\,J1149, provide a helpful constraint which the modelers have used to determine the position of the critical curve near Sp1149. These bright knots, as well as additional constraints provided by the multiple appearances of SN Refsdal, make MACS\,J1149 one of the most well-constrained galaxy clusters. 
MACS\,J1149 is part of the Hubble Frontier Fields \citep[HFF;][]{Lotz_2017} program, and several different teams have created mass models of the cluster using different techniques. In our companion paper (Williams et al., submitted), we compare the magnification predictions at the positions of the \ion{H}{2} regions in Sp1149 from ten different models of the MACS\,J1149 cluster.

In Section~\ref{sec:obs} we describe the OSIRIS and MOSFIRE observations and data reduction. Section~\ref{sec:measurements} details our Bayesian methods for inferring the luminosity and velocity dispersion of the \ion{H}{2} regions in Sp1149. In Section~\ref{sec:calibration}, we describe our new calibration of the $L-\sigma$ relation at $z\approx 0$. We present our results and discuss their implications in Section~\ref{sec:results}, and our conclusions are summarized in Section~\ref{sec:conclusion}.

Throughout this paper, we assume a flat $\Lambda$-cold-dark-matter cosmology with matter density parameter $\Omega_{m}=0.287$ and Hubble parameter H$_0=69.3$~km~s$^{-1}$~Mpc$^{-1}$. All dates and times are reported in UTC. 
\begin{figure}
    \centering
    \includegraphics[width=\linewidth]{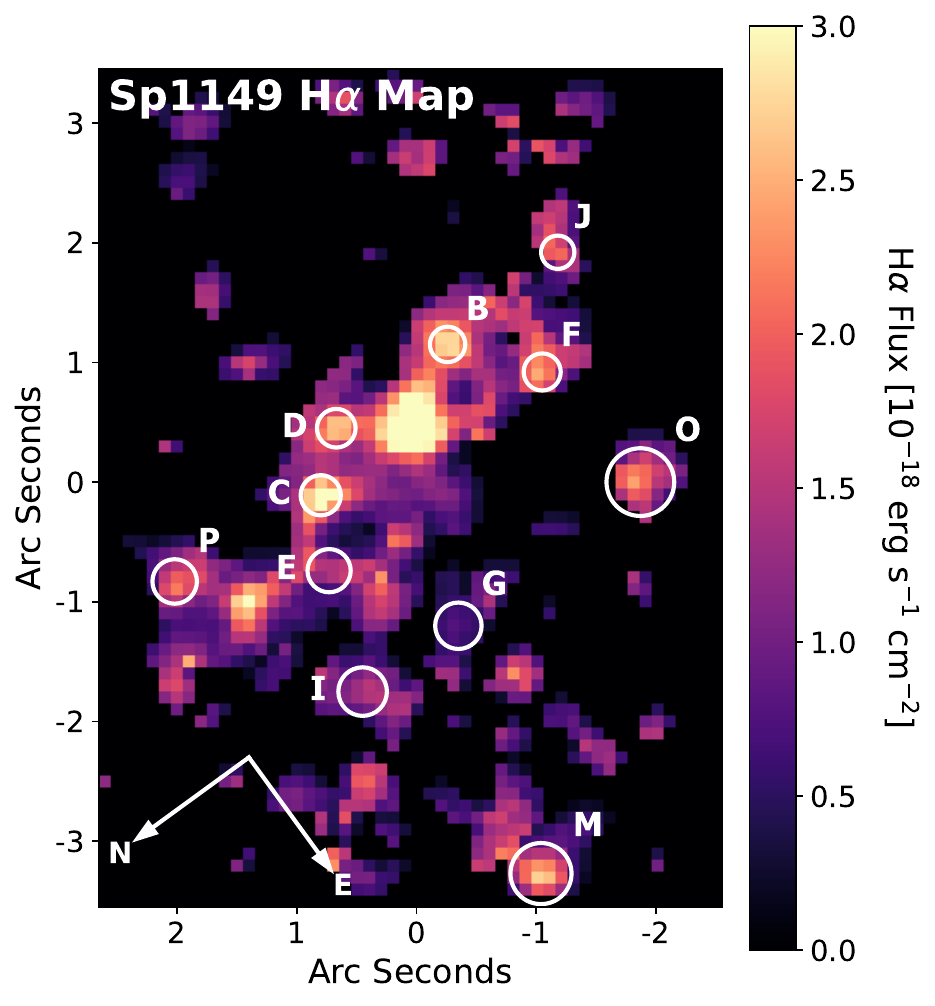}
    \caption{OSIRIS H$\alpha$ emission-line intensity map of Image 1.1 of Sp1149. The $r=500$~pc extraction apertures for the 11 \ion{H}{2} regions used in our analysis are shown as white circles. The \ion{H}{2} region labels are the same as in Figure~\ref{fig:macsj1149}.}
    \label{fig:osiris_HAmap}
\end{figure}

\clearpage

\section{Observations and Data Reduction}\label{sec:obs}
\subsection{OSIRIS}\label{subsec:osiris_obs}

We use existing OSIRIS IFU spectroscopy of Image 1.1 of Sp1149 in the narrow band Hn3 filter (15,940--16,760~\AA; resolution $R\equiv \lambda/\Delta\lambda \approx 3400$) to extract the H$\alpha$ emission-line spectra of the giant \ion{H}{2} regions \citep{Yuan_2011}. The AO system provided a spatial resolution of 0.1$\arcsec$, which corresponds to a typical physical resolution of 300~pc in the source plane. Since the H$\beta$ emission line is not within the spectral range of the OSIRIS data, we rely on Balmer-decrement measurements from MOSFIRE spectroscopy of Sp1149 to correct the OSIRIS H$\alpha$ flux measurements for extinction due to dust (see Section \ref{subsec:MOSFIRE}). Therefore, we only extract those \ion{H}{2} regions from the OSIRIS data that were also observed using MOSFIRE. We extract the spectra of 11 \ion{H}{2} regions from the fully reduced OSIRIS datacube provided by the authors of \cite{Yuan_2011}. 

We extract the one-dimensional (1D) spectrum of each \ion{H}{2} region using a circular aperture with radius $r=500$~pc in the source plane. Using magnification predictions from the {\tt GLAFIC} model of the MACS\,J1149 cluster, we determine the aperture size in the image plane at the position of each \ion{H}{2} region. The aperture radii range from 0.14$\arcsec$ to 0.28$\arcsec$ in the image plane. Figure \ref{fig:osiris_HAmap} shows the H$\alpha$ emission-line intensity map of Image 1.1 of Sp1149, as well as the locations and extraction apertures of the 11 \ion{H}{2} regions we use in our analysis. To test whether the choice of aperture size impacts our final results, we also extract the spectra of the \ion{H}{2} regions using $r=400$~pc and $r=600$~pc circular apertures. We find that these choices of aperture size do not significantly impact our results (see Section \ref{subsec:systematics}). 

We correct the extracted 1D spectra for stellar continuum absorption using {\tt pPXF} \citep[Penalized Pixel-Fitting; ][]{Cappellari_2022} with the MILES stellar library \citep{Sanchez_2006, Falcon_2011}. {\tt pPXF} uses a maximum penalized likelihood method to create a model for the stellar continuum. We subtract {\tt pPXF}'s stellar continuum model for each \ion{H}{2} region from our data to obtain spectra of the nebular emission lines.

\subsection{MOSFIRE}\label{subsec:MOSFIRE}

\subsubsection{MOSFIRE Observations}\label{subsubsec:MOSFIRE_obs}
We observed Sp1149 over two half-nights (February 25 and 26, 2020; hereinafter ``nights 1 and 2") using the MOSFIRE instrument on the Keck-I 10~m telescope \citep{McLean}, as part of NASA-Keck program 75\/2020A\_N110. We used the near-infrared (NIR) \textit{J} and \textit{H} filters, which respectively have the H$\beta$ and H$\alpha$ Balmer lines within their spectral coverage at the redshift of Sp1149 ($z=1.4888\pm0.0011$; \citealt{Teodoro_2018}). 12 distinct \ion{H}{2} regions were targeted in the largest image of Sp1149, labeled as Image 1.1 by \cite{Treu_2016}. The nuclear region (labeled Region A) was observed in both Image 1.1 and Image 1.3, the latter being the image farthest from the critical curve and having the smallest magnification. One \ion{H}{2} region (P) was observed in Image 1.3 but not in Image 1.1. In total, we observed 15 images of 13 different \ion{H}{2} regions. The right panel of Figure \ref{fig:macsj1149} shows the positions of the \ion{H}{2} regions in Images 1.1 and 1.3. 

We acquired the spectra using six MOSFIRE slit masks. For each mask, we selected a position angle so that two or more \ion{H}{2} regions in Image 1.1 were centered on a single slit. For two of the mask configurations, we positioned a second slit on an additional \ion{H}{2} region in Image 1.3. The slit width was 0.7$\arcsec$ and the exposures were taken using an ABAB dither pattern in the \textit{J} and \textit{H} filters. Each integration was 120~s, and the total number of exposures varied on each mask owing to time constraints. Table \ref{tab:masks} shows the \ion{H}{2} regions observed on each mask and the total exposure times. 

On November 16, 2022, we obtained follow-up \textit{H}-band MOSFIRE spectroscopy (PI: A. Filippenko) of \ion{H}{2} Region I in Image 1.1 and Image 1.2 of Sp1149 using one slit mask. The eight exposures (120~s each) were taken using an ABBA dither pattern, for a total exposure time of 16~min. The slit width was 0.7$\arcsec$. 

\begin{table}[ht]
\centering
\ra{1.4}
\caption{MOSFIRE Slit Masks$^a$} 

\begin{tabular}{@{}llll@{}}


&Mask&HII Regions&Exposure Time (\textit{J}, \textit{H})\\
\midrule
\textbf{Feb 25}&1 & 1.1: I, M &28 min, 16 min  \\
\multicolumn{1}{c}{\textbf{2020}} &2 &1.1: A, G &28 min, 16 min\\
&& 1.3: A \\
&3 & 1.1: E, K &8 min, 8 min\\
\midrule
\textbf{Feb 26}&4 & 1.1: B, O &24 min, 12 min\\
\multicolumn{1}{c}{\textbf{2020}} &5 &1.1: C, D, E &36 min, 12 min\\
&&1.3: P\\
&6 &1.1: F, J&28 min, 12 min\\
\midrule
\textbf{Nov 16} & 7 & 1.1: I & N/A, 16 min\\
\multicolumn{1}{c}{\textbf{2022}} & & 1.2: I & \\
\bottomrule
\end{tabular}
{\footnotesize $a$:The seven MOSFIRE slit masks we used for our observations. 0.7$\arcsec$ slits were used in all masks except mask 2, which had 1.0$\arcsec$ slits. We observed a total of 15 images of 13 distinct \ion{H}{2} regions in Sp1149.}
\label{tab:masks}
\end{table}

\begin{figure*}[ht]
    \centering
    \includegraphics[width=1\linewidth]{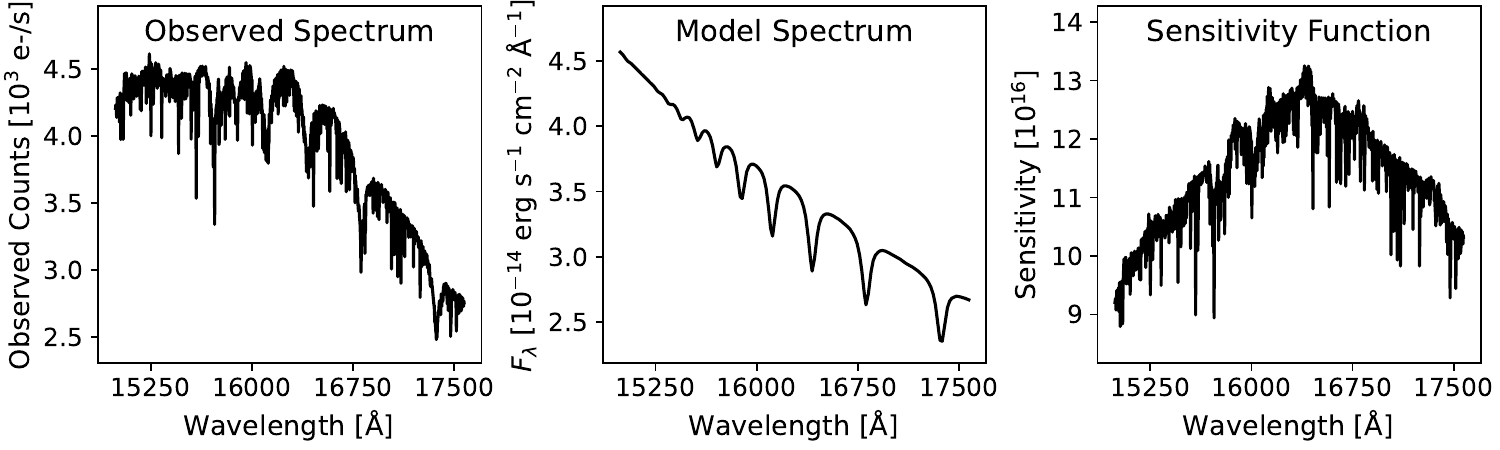}
    \caption{\textit{Left}: The observed standard-star spectrum in the \textit{H} band from the night-one narrow (0.7$\arcsec$) slit observations of HIP~56736. \textit{Middle}: The expected spectrum of the star, calculated by renormalizing a synthetic Vega spectrum to the magnitude of the the standard. \textit{Right}: The relative response function, calculated by dividing the observed spectrum by the model spectrum. We use the sensitivity function to calibrate our science data for the relative response of the detector and correct for telluric absorption.}
    \label{fig:relative_fluxcal}
\end{figure*}

\subsubsection{MOSFIRE Data Reduction}\label{subsubsec:DRP}
We reduced the data using the MOSFIRE {\tt Data Reduction Pipeline} \citep[{\tt DRP};][]{Konidaris_2019}. The pipeline uses dome-flat exposures to trace the slit edges and create a pixel-flat correction. Next, an argon comparison-lamp observation was used to generate a wavelength solution. In the following step, the pipeline combines all of the science frames in the ``A" dither position and all those in the ``B" position, then subtracts B from A to obtain background-subtracted, coadded science spectra. Finally, the {\tt DRP} combines the wavelength solution with the background-subtracted spectra to produce a spatially rectified 2D spectrum for each slit. The {\tt DRP} also produces a signal-to-noise ratio (S/N) map for every slit. The S/N is computed for each pixel by adding together the signal in each frame and then dividing by the square root of the variance between the frames.

At the beginning of each half-night, we obtained an observation of a telluric standard star at an airmass comparable to that of the MACS\,J1149 field: HIP~56736 on night 1 and HIP~55627 on night 2 \citep{standard_catalog}. We also attempted to obtain an additional telluric standard observation at the end of each night, but were prevented by time constraints on the first night and malfunction of the guiding system during the second night. We use the standard-star observations to measure the relative response function along the spectral axis and correct for telluric absorption.

We extract the spectrum of each standard star, using a boxcar aperture whose width is twice the full width at half-maximum intensity (FWHM) of the star's flux profile along the spatial axis. We observed A0~V-type standard stars, so we use the {\tt synphot} \citep[][]{synphot} software to generate a synthetic Vega spectrum scaled to the standard stars' respective magnitudes as the calibrated model. We then calculate the relative response function  by dividing the observed spectrum by the model spectrum. An example of a relative response function is shown in Figure \ref{fig:relative_fluxcal}.

\begin{figure}
    \centering
    \includegraphics[width=\linewidth]{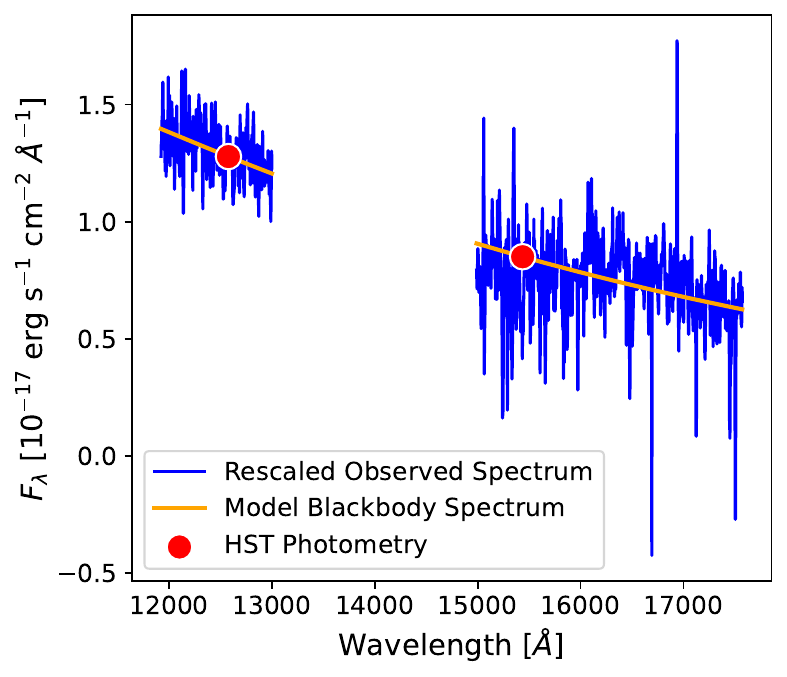}
    \caption{Observed MOSFIRE spectrum of a field star on mask 4 (blue line), renormalized to match {\it HST} WFC3 F125W and F160W photometry (red points). The renormalization factor is computed by comparing the observed spectrum to a model blackbody spectrum (orange line). We use the renormalization factor to determine the absolute flux calibration for each MOSFIRE mask.}
    \label{fig:absolute_fluxcal}
\end{figure}

To obtain an absolute flux calibration for each set of observations with each mask, we included field stars on each mask positioned on 0.7$\arcsec$-wide slits (1.0$\arcsec$ for mask 2). We extract a 1D spectrum of each star using a boxcar aperture with a width equal to the FWHM of the spatial profile of the star to match the extraction apertures of our science data, and multiply the extracted spectrum by the relative response function. Since the spectrum of each field star was obtained using a slit width that matches that employed for the science targets,  our absolute flux calibration procedure accounts for slit losses from a point source. To calculate the expected flux of the stars, we use \textit{HST} photometry in the WFC3-IR F125W (wide {\it J}) and WFC3-IR F160W ({\it H}) filters, which provide the closest matches to the central wavelengths of the MOSFIRE \textit{J} and \textit{H} filters. We fit a blackbody spectrum to the photometry, and then divide our telluric-corrected spectrum by the model blackbody spectrum and take the median value to compute the absolute flux normalization factor.  Figure \ref{fig:absolute_fluxcal} shows the renormalized spectrum of one of the slit stars with the \textit{HST} photometry. 

We next calibrate the science data  by multiplying the spectra by both the relative response function and the absolute flux normalization factor. For the masks on which we observed more than one field star, the flux-calibration factors calculated for each star agree within 15\%. On mask 3, we found that the flux-calibration factors calculated for two different field stars differed by more than 50\%, likely due to a mask alignment issue. Therefore, \ion{H}{2} Region K was removed from the sample and we used mask 5 for our measurements of Region E.

At the redshift of Sp1149 ($z=1.49$), the H$\beta$ emission line lies in the \textit{J} band ($\lambda_{\rm obs,H\beta} = 12,099$~\AA) and the H$\alpha$ emission line falls in the \textit{H} band ($\lambda_{\rm obs,H\alpha} = 16,333$~\AA). Since we use the observed flux ratio of H$\alpha$/H$\beta$ to estimate the extinction due to dust, our inference about the extinction is sensitive to the relative \textit{J} and \textit{H} flux calibration. To estimate the uncertainty in their relative calibration, we use the three masks (masks 1, 2, and 4) that each have two usable stars. For each \ion{H}{2} region on these masks, we compare the measured flux ratio of H$\alpha$/H$\beta$ using the absolute flux calibration factors computed for each field star. We find that the fractional difference between the flux ratios computed for each star is $<5\%$ for each \ion{H}{2} region. Thus, we adopt an uncertainty estimate of 5\% for the relative flux calibration. 

We extract each \ion{H}{2} region using a boxcar method with the aperture width set as the FWHM of the field star on the same mask and filter. The relative and absolute flux calibration is applied to each extracted spectrum. We correct the extracted spectra for stellar continuum absorption using {\tt pPXF} (see Section~\ref{subsec:osiris_obs}) The spectra are corrected for the foreground extinction due to dust in the Milky Way of 0.016~mag in the \textit{J}-band and 0.010~mag in the \textit{H}-band \citep{Shlafly_2011}.

\subsection{LBT Observations and Data Reduction}
On February 13, 2021, we obtained follow-up \textit{H}-band spectroscopy of \ion{H}{2} Regions G and M with the LUCI \citep[LBT Utility Camera in the Infrared,][]{LUCI_instrument} spectrographs on the Large Binocular Telescope \citep[LBT;][]{LucI}. We obtained a 1~hr \textit{H}-band integration with the \ion{H}{2} regions centered on the LUCI long slit with a slit width of 0.5$\arcsec$. Each exposure was 180~s long, and we used an ABBA dither pattern. The G210 grating, which has a spectral resolution of $R \approx 5900$ in the \textit{H} band, was used for the observations and the seeing FWHM was 0.5$\arcsec$. 

The LUCI data were reduced using {\tt flame}, an IDL pipeline designed for LUCI spectroscopy \citep[][]{flame}. The pipeline uses dome-flat exposures to generate a bad-pixel mask and then measure a flatfield correction, which it applies. The package next uses bright OH night-sky lines to trace the edges of the slits and perform spatial rectification. The step that follows is to use the sky lines to calculate the wavelength solution. Finally, the pipeline subtracts the science frames in the ``B" dither position from those in the ``A" dither position to produce a background-subtracted science spectrum.

Immediately following our observations of Sp1149, we acquired spectra of the telluric standard star HIP 62745 on the same long-slit mask. We use our spectra of this star to compute the relative flux calibration, correct for telluric absorption, and compute the absolute flux calibration factor using the same methods as we did for the MOSFIRE observations. After applying these corrections to the long-slit spectrum, we extract \ion{H}{2} Regions G and M using a boxcar extraction with the aperture width set as the observed FWHM of HIP 62745. 

\subsection{Consistency Across Observations}
To evaluate any systematic uncertainties affecting our observations and data reduction, we measure the H$\alpha$ fluxes and emission-line widths of \ion{H}{2} Regions G and M from the reduced OSIRIS, MOSFIRE, and LUCI spectra. For an accurate comparison, we convolve the OSIRIS AO-assisted IFU datacube to match the $\sim 0.7$\arcsec \ natural seeing of the LUCI and MOSFIRE data and extract the 1D spectra of Regions G and M from the OSIRIS IFU datacube using a square 0.7$\arcsec$ aperture. 

We use a nonlinear least-squares fitting algorithm in the {\tt scipy} \citep{2020SciPy-NMeth} package to model the H$\alpha$ emission line in each spectrum. To accurately compare the line widths measured from different instruments, we subtract the instrumental profile in quadrature from the observed line width. The measured H$\alpha$ fluxes and line fluxes are listed in Table~\ref{tab:data_consistency}. We find that the fluxes and line widths we measure from the OSIRIS, MOSFIRE, and LUCI spectroscopy are consistent  within their 1$\sigma$ uncertainties. 

\begin{table}[ht]

\ra{1.3}
\caption{Observational Consistency Check$^{a}$} 
\centering
\begin{tabular}{@{}llll@{}}
\toprule
 \multicolumn{4}{c}{\textbf{H$\alpha$ Flux (10$^{-17}$ erg s$^{-1}$ cm$^{-2}$)}}\\
 \ion{H}{2} Region& OSIRIS &MOSFIRE&LUCI\\

 Region G & 4.0 $\pm$ 0.5 & 3.8 $\pm$ 0.5 &4.2 $\pm$ 1.0 \\
 Region M & 7.9 $\pm$ 0.8 & 7.6 $\pm$ 0.4 & 7.1 $\pm$ 1.0\\

 \multicolumn{4}{c}{\textbf{H$\alpha$ Line Width (km s$^{-1}$)}}\\
 \ion{H}{2} Region& OSIRIS &MOSFIRE&LUCI\\

Region G &17.6 $\pm$ 14.2 & 21.8 $\pm$ 15.2 & 13.3 $\pm$ 10.4\\
Region M & 14.9 $\pm$ 12.9 & 14.9 $\pm$ 6.5 & 16.0 $\pm$ 5.6 \\
\bottomrule
\end{tabular}
{\footnotesize $^a$Emission-line flux and width measurements from the OSIRIS, MOSFIRE, and LUCI spectroscopy of \ion{H}{2} Regions G and M. The reported line width has the respective instrumental profile subtracted in quadrature for each observation. We convolve the OSIRIS data to match the $\sim0.7$\arcsec \ natural seeing of the LUCI and MOSFIRE spectroscopy.}

\label{tab:data_consistency}
\end{table}

\section{Measurements}\label{sec:measurements}
With fully reduced and calibrated spectra, we apply Bayesian methods to infer the intrinsic Balmer luminosity and velocity dispersion of each \ion{H}{2} region. Our analysis is a two-step process. First, we use the MOSFIRE spectra, which include both H$\alpha$ and H$\beta$, to infer the rest-frame {\it V}-band dust extinction in magnitudes $A_V$ for each \ion{H}{2} region. Next, we use the inferred $A_V$ values to correct the OSIRIS spectra for extinction and measure the intrinsic H$\alpha$ luminosity and velocity dispersion of each \ion{H}{2} region from the OSIRIS H$\alpha$ emission-line profiles.  

\subsection{Emission-Line Fluxes and Extinction from MOSFIRE Data}\label{subsec:Av_measurements}
We measure the H$\beta$, [\ion{O}{3}], H$\alpha$, and [\ion{N}{2}] emission-line fluxes in the MOSFIRE spectra. The Balmer decrement is used to infer the internal galaxy dust extinction for each \ion{H}{2} region with a Bayesian inference method, and we use the fluxes of the four emission lines we measured to place the \ion{H}{2} regions on the Baldwin-Phillips-Terlevich (BPT) diagram \citep{Baldwin_1981}.  We model the emission-line profiles using a set of nine parameters: the fluxes of the H$\alpha$, [\ion{N}{2}], and [\ion{O}{3}] emission lines ($F_{{\rm H}\alpha}$, $F_{{\rm [N\,II]}}$, and $F_{\rm [O\,III]}$, respectively), the intrinsic velocity dispersion ($\sigma$), the rest-frame {\it V}-band extinction due to dust ($A_V$), the observer-frame central wavelength of the H$\alpha$ emission line ($\lambda_{\rm obs, H\alpha}$), and the flux density of the continuum adjacent to three emission lines or closely spaced lines ($C_{\rm H\beta}$, $C_{\rm [O\,III]}$, and $C_{\rm H\alpha + [N\,II]}$). An uninformative and uniform prior is used for each of these parameters. We include an additional, tenth parameter ($JH$) to account for the relative-flux calibration between the $J$ and $H$ filters. We estimate a 5\% uncertainty in the relative-flux calibration (see Section~\ref{subsubsec:DRP}), so the prior for $JH$ is a normal distribution centered at 1 with a standard deviation of 0.05. All of the parameters used in our model and priors on their values are listed in Table \ref{tab:extinction_priors}.

\begin{table}
\ra{1.3}
\centering
\caption{MOSFIRE Extinction Model Priors$^a$}
\begin{tabular}{@{}cll@{}}

Parameter(s) & Prior & Bounds\\
\midrule
$F_{{\rm H}\alpha}$, $F_{{\rm [N\,II]}}$, & Uniform & [0, 50] $10^{-17}$~erg~s$^{-1}$~cm$^{-2}$ \\
\multicolumn{1}{c}{$F_{\rm [O\,III]}$} &&\\

$\sigma$ & Uniform &[0, 250] km s$^{-1}$\\

$A_V$ & Uniform &[0, 10] mag\\

$\lambda_{\rm obs, H\alpha}$ & Uniform&[16320, 16355] \AA\\

$C_{\rm H\beta}$, $C_{\rm [O\,III]}$, & Uniform &[-1, 1] $10^{-17}$ erg s$^{-1}$ cm$^{-2}$ \AA$^{-1}$ \\

\multicolumn{1}{c}{$C_{\rm H\alpha + [N\,II]}$} && \\

$JH$ & Normal & $\mu=1.00$, $\sigma=0.05$\\

\bottomrule
\end{tabular}
{\footnotesize $^a$Prior constraints placed on each of the parameters used in our model for the MOSFIRE spectroscopy. We use Uniform priors with large parameter spaces. $JH$ allows us to marginalize over the uncertainty in the flux calibrations of the {\it J} and {\it H} bands.}
\label{tab:extinction_priors}
\end{table}

For a given set of values of the model parameters, we can produce model Gaussian profiles of the H$\beta$, [\ion{O}{3}], H$\alpha$, and [\ion{N}{2}] emission lines and compare these to the observed profiles to compute the likelihood of the data.  
To compute the model profile of each emission line, we need to convert the parameter values to the observed flux, dispersion, and central wavelength of the line.

The fluxes of H$\alpha$, [\ion{N}{2}], and [\ion{O}{3}] are model parameters. We compute the flux of the H$\beta$ emission line $F_{\rm H\beta}$ from $F_{{\rm H}\alpha}$ and $A_V$ parameters given the H$\alpha$/H$\beta$ ratio expected from Case~B recombination \citep{Osterbrock}. 
The \citet{Cardelli_1989} extinction law is used with $R_V=3.10$ to calculate the extinction at the wavelengths of H$\beta$ and H$\alpha$. Our analysis is repeated using  $R_V =2.51$ from the MOSFIRE Deep Evolution Field (MOSDEF) survey of galaxies at $1.4\lesssim z \lesssim 2.6$ \citep{Reddy_2015}, and we find that the choice of $R_V$ does not impact our results (see Section \ref{subsec:systematics}). We multiply the value of the flux of H$\beta$ we calculate by the $JH$ parameter to account for the 5\% uncertainty in the relative flux calibration between the $J$ and $H$ filters.

The total velocity dispersion consists of four components,
\begin{equation}\label{equation:dispersion}
    \sigma_{\rm tot}^2 = \sigma^2 + \sigma_{\rm ins}^2 + \sigma_{\rm th}^2 + \sigma_{\rm fs}^2,
\end{equation}
where $\sigma_{\rm ins}$ is the instrumental broadening, $\sigma_{\rm th}$ is the thermal broadening, and $\sigma_{\rm fs}$ is the fine-structure broadening. The instrumental broadening of the MOSFIRE spectrograph\footnote{http://www2.keck.hawaii.edu/inst/mosfire/grating} is 39~km~s$^{-1}$ at the observer-frame wavelength of H$\beta$ and 35~km~s$^{-1}$ at the observer-frame wavelength of H$\alpha$. We adopt values of $\sigma_{fs}^2(H\alpha) = 10.233$~km$^2$~s$^{-2}$  and $\sigma_{fs}^2(H\beta) = 5.767$~km$^2$~s$^{-2}$  from \cite{garcia}. Fine-structure broadening is not significant for metal lines, so it is not included in the calibration for [\ion{O}{3}] and [\ion{N}{2}]. The thermal broadening is computed from the following expression assuming a Maxwellian velocity distribution for the hydrogen, oxygen, and nitrogen ions:

\begin{equation}
    \sigma_{\rm th} = \sqrt{\frac{kT_e}{m}}\, ,
\end{equation}
\noindent
where $k$ is the Boltzmann constant, $m$ is the mass of the ion, and $T_e$ is the electron temperature. We adopt a typical value of $T_e = 8000$~K for the \ion{H}{2} regions in our sample, which corresponds to $\sigma_{\rm th} = 8.1$~km~s$^{-1}$ for the hydrogen ion.

\begin{table}[]
    \centering
    \ra{1.3}
    \setlength{\tabcolsep}{10pt}
    \caption{\ion{H}{2} Region Properties$^a$}
    \begin{tabular}{clll}

    Region & $A_V$ (mag) & $\mu$ (GLAFIC) & Seeing\\
    \midrule
    B & $0.15^{+0.22}_{-0.11}$ & $6.5^{+0.3}_{-0.3}$ & 339 pc\\
    C & $0.10^{+0.15}_{-0.08}$ & $8.4^{+0.3}_{-0.4}$ & 298 pc\\
    D & $1.12^{+0.48}_{-0.45}$ & $7.8^{+0.3}_{-0.4}$ & 309 pc\\
    E & $0.25^{+0.34}_{-0.18}$ & $9.7^{+0.4}_{-0.5}$ & 277 pc\\
    F & $1.07^{+0.94}_{-0.67}$ & $7.2^{+0.3}_{-0.5}$ & 322 pc \\
    G & $0.76^{+0.72}_{-0.49}$ & $11.3^{+0.7}_{-0.7}$ & 257 pc\\
    I & $1.43^{+1.33}_{-0.82}$ & $12.3^{+0.7}_{-0.7}$ & 246 pc \\
    J & $1.85^{+1.14}_{-0.92}$ & $5.9^{+0.3}_{-0.3}$ & 355 pc\\
    M & $1.38^{+0.70}_{-0.60}$ & $19.5^{+1.7}_{-1.7}$ & 195 pc\\
    O & $0.23^{+0.35}_{-0.17}$ & $24.1^{+3.1}_{-3.3}$ & 176 pc\\
    P & $0.33^{+0.49}_{-0.24}$ & $3.1^{+0.1}_{-0.1}$ & 490 pc\\
    \bottomrule
    \end{tabular}
    {\footnotesize $^a$Dust extinction measurements and GLAFIC magnification predictions for each of the \ion{H}{2} regions used in our analysis. The value of $A_V$ is the median of the posterior on the MCMC fit to our MOSFIRE data, and the magnification is the median value of 100 MCMC realizations of the GLAFIC model of the MACS\,J1149 cluster. The reported uncertainties are the 16th and 84th percentiles.}
    \label{tab:hii_properties}
\end{table}

The observer-frame central wavelength of H$\alpha$ is given by the model parameter $\lambda_{\rm obs,H\alpha}$, and we compute the observer-frame central wavelength of the other three emission lines by multiplying $\lambda_{\rm obs,H\alpha}$ by the ratio of the rest wavelengths of H$\beta$, [\ion{O}{3}], and [\ion{N}{2}] to H$\alpha$. 

After obtaining the flux, dispersion, and central wavelength of each emission line from a given set of model parameter values, we can calculate the model flux density $F_{\rm mod}$ as a function of wavelength,

\begin{equation}\label{equation:line_model}
    F_{\rm mod}(\lambda)=C + \frac{F}{\sqrt{2\pi}\sigma_{\rm tot}} e^{-\frac{(\lambda-\lambda_{\rm obs})^2}{2\sigma_{\rm tot}^2}}\, .
\end{equation}




\noindent
To find the best-fit set of model parameters, we create a likelihood function which compares the MOSFIRE spectrum with the model spectrum. The likelihood function for the observations of each \ion{H}{2} region is given by
\begin{equation}\label{equation:likelihood}
    L({F}_{\rm obs} \vert {x})=\prod_{i=1}^n 
    \frac{1}{\sqrt{2\pi}\sigma_{\rm tot,i}} e^{-\frac{1}{2 \sigma_{\rm tot,i}^2}(F_{\rm mod,i}({x})-F_{\rm obs,i})^2}\, ,
\end{equation}
where $F_{\rm obs,i}$ is the observed flux density of each data point, $F_{\rm mod,i}$ is the model flux density calculated at the observed wavelength of each data point, and ${x}$ represents the set of model parameters. We fit the MOSFIRE spectra within $\pm 20\sigma$ ($\sim 40$~\AA) around the wavelengths of the emission lines.

We sample from the posterior probability distribution using a Markov Chain Monte Carlo (MCMC) algorithm implemented by the {\tt Python} package {\tt pymc} \citep{pymc}. The No-U-Turn Sampler \citep[NUTS;][]{hoffman2011nouturn} is used to generate sample values for all ten parameters at each step. NUTS is a Hamiltonian Monte Carlo (HMC) sampling method that chooses its steps based on the first-order gradient of the likelihood function and uses a recursive algorithm to select a set of likely candidate points. We sample the posterior 10,000 times after 5000 burn-in steps. 

The 16th, median, and 84th percentiles are used to define our 1$\sigma$ confidence intervals. Best-fit emission-line profiles are shown in Figure~\ref{fig:mosfire_profiles}. We use the posterior on $A_V$ to correct the OSIRIS H$\alpha$ fluxes for dust extinction and calculate the luminosity of the \ion{H}{2} regions, and we use the posteriors on the fluxes of the strong nebular emission lines to constrain the locations of the \ion{H}{2} regions on the BPT diagram (see Figure~\ref{fig:BPT}). The best-fit extinction measurements are given in Table~\ref{tab:hii_properties}.

We have already corrected for the foreground Milky Way extinction, and extinction through the MACS\,J1149 cluster is expected to be negligible, given the extreme temperatures of the intracluster gas. Dust lanes in the cluster-member galaxy around which the images of SN Refsdal form an Einstein cross could contribute to the extinction, but would likely only affect Region O (see Figure~\ref{fig:macsj1149}). We expect, therefore, that the Balmer decrements we measured are due to extinction in Sp1149. We find that the average extinction among the 11 \ion{H}{2} regions is $A_V=0.79\pm0.53$~mag, consistent with a previous measurement of the global extinction in Sp1149 from {\it HST} grism spectroscopy \citep[$A_V=0.72\pm0.23$~mag;][]{Wang_2017}. The dispersion among the extinction measurements in the 11 \ion{H}{2} regions in Sp1149 ($0.53$~mag) is typical for \ion{H}{2} regions in the nearby galaxies that form our local sample for the $L-\sigma$ calibration (see Section~\ref{sec:calibration}). The extinction dispersions among the \ion{H}{2} regions in those nine local galaxies range from 0.26~mag to 0.56~mag.

\subsection{OSIRIS $L$ and $\sigma$ Measurements}\label{subsec:osiris_measurements}
We use the H$\alpha$ flux measured from the OSIRIS IFU spectra together with the extinction measured from the MOSFIRE spectroscopy to infer the intrinsic H$\alpha$ luminosity of each \ion{H}{2} region using Bayesian inference. The OSIRIS observations are used to infer the velocity dispersion for each \ion{H}{2} region. We model the H$\alpha$ profiles with a set of four parameters: the logarithm of the intrinsic luminosity of H$\alpha$ ($\log_{10}({\rm L_{H\alpha}})$), the intrinsic velocity dispersion ($\sigma$), the observed central wavelength of the H$\alpha$ emission line ($\lambda_{\rm o,H\alpha}$), and the continuum flux density near the H$\alpha$ line ($C_{\rm H\alpha}$). An uninformative and uniform prior is again applied. We include an additional (fifth) parameter ($f$) which allows us to marginalize over the 10\% uncertainty in the absolute-flux calibration of the OSIRIS spectroscopy \citep{Yuan_2011}. The prior for $f$ is a normal distribution centered at 1, with a standard deviation of 0.10. All of the parameters used in our model and their priors are shown in Table \ref{tab:osiris_priors}. To account for the extinction due to dust and incorporate the uncertainty in that measurement, we randomly sample from the posterior on $A_V$ from the fit to our MOSFIRE data for each \ion{H}{2} region (see Section~\ref{subsec:Av_measurements}). 

\begin{table}
\ra{1.3}
\centering
\caption{OSIRIS $L-\sigma$ Model Priors$^a$}
\begin{tabular}{@{}lll@{}}
Parameter&Prior&Bounds\\
\midrule
$\log_{10}({\rm L_{H\alpha}})$ & Uniform & [36, 44] erg s$^{-1}$\\

$\sigma$ & Uniform &[0, 250] km s$^{-1}$\\

$\lambda_{\rm obs,H\alpha}$ & Uniform & [16320, 16355] \AA\\

$C_{\rm H\alpha}$ & Uniform &[-1, 1] $10^{-17}$ erg s$^{-1}$ cm$^{-2}$ \AA$^{-1}$ \\
$f$ & Normal & $\mu=1.00$, $\sigma=0.10$\\

\bottomrule
\end{tabular}
{\footnotesize $^a$Prior constraints placed on each of the parameters used in the model for the OSIRIS data. We use uniform priors with large parameter spaces. The 10\% uncertainty in the absolute flux calibration of the OSIRIS data is represented by $f$.}
\label{tab:osiris_priors}
\end{table}

\begin{table}[]
\centering
\ra{1.3}
\setlength{\tabcolsep}{16pt}
\caption{Sp1149 $L$ and $\sigma$ Measurements$^a$}
\begin{tabular}{@{}ccl@{}}
\ion{H}{2} Region & $\log(L_{{\rm H}\alpha})$ [erg s$^{-1}$] & $\sigma$ [km s$^{-1}$]\\
\midrule
B & $40.68^{+0.12}_{-0.11}$ & $44.46^{+20.23}_{-19.83}$\\
C & $40.72^{+0.09}_{-0.09}$ & $60.28^{+12.89}_{-13.39}$\\
D & $40.97^{+0.22}_{-0.21}$ & $61.61^{+28.37}_{-21.10}$\\
E & $40.55^{+0.14}_{-0.12}$ & $36.99^{+16.55}_{-18.08}$\\
F & $40.71^{+0.34}_{-0.28}$ & $41.98^{+32.64}_{-27.44}$\\
G & $40.49^{+0.24}_{-0.18}$ & $7.60^{+8.46}_{-5.37}$\\
I & $41.00^{+0.45}_{-0.28}$ & $4.00^{+4.49}_{-2.90}$\\
J & $41.21^{+0.39}_{-0.33}$ & $88.10^{+27.32}_{-21.81}$\\
M & $40.78^{+0.24}_{-0.21}$ & $36.93^{+9.78}_{-10.07}$\\
O & $40.45^{+0.14}_{-0.12}$ & $33.84^{+8.96}_{-9.83}$\\
P & $40.72^{+0.18}_{-0.13}$ & $56.13^{+13.21}_{-13.80}$\\
\bottomrule
\end{tabular}
{\footnotesize $^a$Measured values of the logarithm of the intrinsic H$\alpha$ luminosity $\log(L_{{\rm H\alpha}})$ and the intrinsic velocity dispersion $\sigma$ for the \ion{H}{2} regions in Sp1149 ($z=1.49$).}
\label{tab:measurements}
\end{table}

To account for gravitational lensing magnification and convert between the observed flux and intrinsic luminosity of each \ion{H}{2} region, we use predictions for the magnification from the {\tt GLAFIC} simply parameterized model of the MACS\,J1149 cluster \citep{Oguri_2010,Kawamata_2016}. The modeling team provides 100 MCMC realizations of their models, which are available for download at the Mikulski Archive for Space Telescopes (MAST) HFF Data Access Page\footnote{https://archive.stsci.edu/pub/hlsp/frontier}. To incorporate the uncertainty associated with the magnification prediction, we randomly sample from the magnification values $\mu$ which correspond to realizations of the lens model for each \ion{H}{2} region. The GLAFIC magnification predictions are listed in Table~\ref{tab:hii_properties}.

Given a set of values for the model parameters, along with the inferred extinction from the {\it J}-band and {\it H}-band  MOSFIRE data and the predicted magnification from the GLAFIC model, we can compute a model of an H$\alpha$ line profile for each \ion{H}{2} region.
We first compute the apparent H$\alpha$ flux from the intrinsic H$\alpha$ luminosity and magnification together with extinction due to dust. We use the value of the magnification from the {\tt GLAFIC} lens model to calculate the magnified H$\alpha$ luminosity $\log_{10}(L_{\rm H\alpha,lens})$,

\begin{equation}
    \log_{10}(L_{\rm H\alpha,lens}) = \log_{10}(L_{H\alpha}) + \log_{10}(\mu).
\end{equation}

Assuming Case B recombination and using the value of $A_V$ that we inferred from  the Balmer decrement observed in the MOSFIRE spectra, we calculate the  luminosity of H$\alpha$ after extinction from dust from $L_{\rm H\alpha,lens}$. For a luminosity distance to Sp1149 of 11~Gpc, we calculate the dust-extinguished flux ($F_{\rm H\alpha}$) from the line luminosity. To account for the 10\% uncertainty in the absolute-flux calibration of the OSIRIS data, we multiply the value of $F_{\rm H\alpha}$ by the value of the model parameter $f$. 

The width of the H$\alpha$ emission-line profile is calculated using Eq.~\ref{equation:dispersion}. We measure the instrumental broadening as a function of wavelength in the OSIRIS Hn3 filter using the observed widths of night-sky lines. The instrumental broadening is 39~km~s$^{-1}$ at the observer-frame wavelength of H$\alpha$. Model parameter $\lambda_{\rm obs,H\alpha}$ specifies the observer-frame central wavelength of H$\alpha$.

From the H$\alpha$ flux, width, and central wavelength along with the flux density of the continuum, we use Eq.~\ref{equation:line_model} to compute a model for the OSIRIS data.
The likelihood of the OSIRIS data is computed using Eq.~\ref{equation:likelihood}. 
We fit the OSIRIS H$\alpha$ profiles within 20~$\sigma$ windows ($\sim 40$~\AA) around the observer-frame central wavelength of H$\alpha$.    

We sample from the parameter space using the same MCMC method as in Section~\ref{subsec:Av_measurements}. The median of the posteriors is adopted as the best-fit value for each parameter and the 16th and 84th percentile values as the 1$\sigma$ uncertainties. The values for the intrinsic H$\alpha$ luminosity and velocity dispersion of each \ion{H}{2} region are shown in Table~\ref{tab:measurements}. The best-fitting emission-line profiles for the OSIRIS data are shown in Figure~\ref{fig:osiris_profiles}.

\section{$L-\sigma$ Relation at Low Redshift}\label{sec:calibration}
Existing calibrations of the $L-\sigma$ relation use high spatial and spectral resolution ($R\gtrsim10,000$) observations of giant \ion{H}{2} regions in nearby ($\lesssim 20$~Mpc) galaxies \citep[e.g.,][]{Chavez_2012,Fernandez_Arenas_2018} to measure the luminosities and velocity dispersions. At these small distances, \ion{H}{2} regions can be extracted with physical aperture widths smaller than $\sim 100$~pc and their luminosities and velocity dispersions can be measured with minimal contamination from diffuse ionized gas (DIG) and other background sources \citep[see Figure A1 of][]{Fernandez_Arenas_2018}. Even with the assistance of magnification by gravitational lensing, the distance of Sp1149 ($D_A \approx 1780$~Mpc) makes the spatial resolution much poorer than these local measurements, and our 500~pc aperture sizes may include contamination from the DIG and contributions from more than one \ion{H}{2} region. Additionally, the spectral resolution of the OSIRIS observations ($R=3250$) is much lower than the extremely high resolution spectroscopy used in existing $L-\sigma$ calibrations ($R \approx 10,000$--20,000). With lower spectral resolution, emission-line widths are more difficult to measure precisely. At lower spectral resolution, inferred line widths from fitting a Gaussian to the line profile can be sensitive to non-Gaussian tails or structure in the line. By simulating emission lines convolved with $R \approx 3000$ and  $R \approx 10,000$ instrumental profiles, we estimate that intrinsic emission-line widths measured with $R \approx 3000$ spectroscopy can differ from those measured with $R \approx 10,000$ by up to $\sim20\%$. Therefore, a calibration of the $L-\sigma$ relation using physical aperture sizes of $\sim 500$~pc and $R\approx 3000$ spectroscopy is necessary to draw an accurate comparison between our measurements of the \ion{H}{2} regions in Sp1149 with the luminosities and velocity dispersions of \ion{H}{2} regions in low-redshift galaxies.

\begin{table}[ht]
\centering
\ra{1.5}
\scriptsize
\caption{Galaxies Used in $L-\sigma$ Calibration$^a$}
\begin{tabular}{@{}lllcc@{}}

Galaxy Name& SN Name & Redshift &Regions & Seeing\\
\midrule
2MASSX\,J01504127 & ASASSN-14hr &0.0336  & 38 & 790 pc\\
2MASSX\,J09202045 & ASASSN-15fr &0.0334 & 23 & 560 pc\\
CGCG\,023-005 & ASASSN-14co & 0.0333 & 50 & 520 pc\\
CGCG\,048-099 & SN2014ey &0.0320 & 35 & 390 pc\\
E153-G020 & SN2012U &0.0197 & 37 & 380 pc\\
E287-G040 & SN2013fy & 0.0302 & 31 & 920 pc\\
E570-G020 & SN2009aa & 0.0273 &48 & 340 pc\\
NGC\,2370 & PSNJ07250 &  0.0184 & 69 & 430 pc\\
SDSS\,J103247 & LSQ13ry & 0.0299& 16 & 400 pc\\
\bottomrule
\end{tabular}
{\footnotesize $^a$Galaxies are selected from the AMUSING survey of SN host galaxies. The H$\beta$ luminosity and velocity dispersion of 347 total \ion{H}{2} regions in these 9 galaxies are measured from MUSE IFU spectroscopy, using the same aperture sizes as  for our analysis of the OSIRIS IFU data.}
\label{tab:muse_galaxies}
\end{table}

We perform a new measurement of the $L-\sigma$ relation at low redshift using existing IFU spectroscopy of galaxies from the All-weather MUse Supernova Integral field Nearby Galaxies (AMUSING) survey \citep{Galbany_2016}, which uses the Multi Unit Spectroscopic Explorer (MUSE) on the Very Large Telescope (VLT) to observe the host galaxies of SNe. The spectral resolving power of MUSE is $R \approx 3000$ \citep{Bacon_2010}, which is comparable to that of OSIRIS. To select galaxies from the AMUSING survey for our calibration of the $L-\sigma$ relation, we apply the following criteria: (i) The galaxy must be a spiral at a relatively small inclination angle similar to that of Sp1149. (ii) There must be  \ion{H}{2} regions that are easily  identified by eye in the MUSE data of the galaxy. The AMUSING survey observes in any conditions to make use of nonoptimal weather at the VLT, so many objects are observed in very poor seeing and therefore the seeing is not sufficient to resolve  \ion{H}{2} regions in a significant fraction of the archival data. We require the physical spatial resolution FWHM to be $< 1$~kpc. (iii) We select galaxies within a redshift range of $0.015 \lesssim z \lesssim 0.040$. The lower bound limits  uncertainties on the galaxy's distance inferred from its redshift, and the upper bound ensures that the spatial resolution is small enough to distinguish between individual \ion{H}{2} regions. 

\begin{figure*}
    \centering
    \includegraphics[width=.9\linewidth]{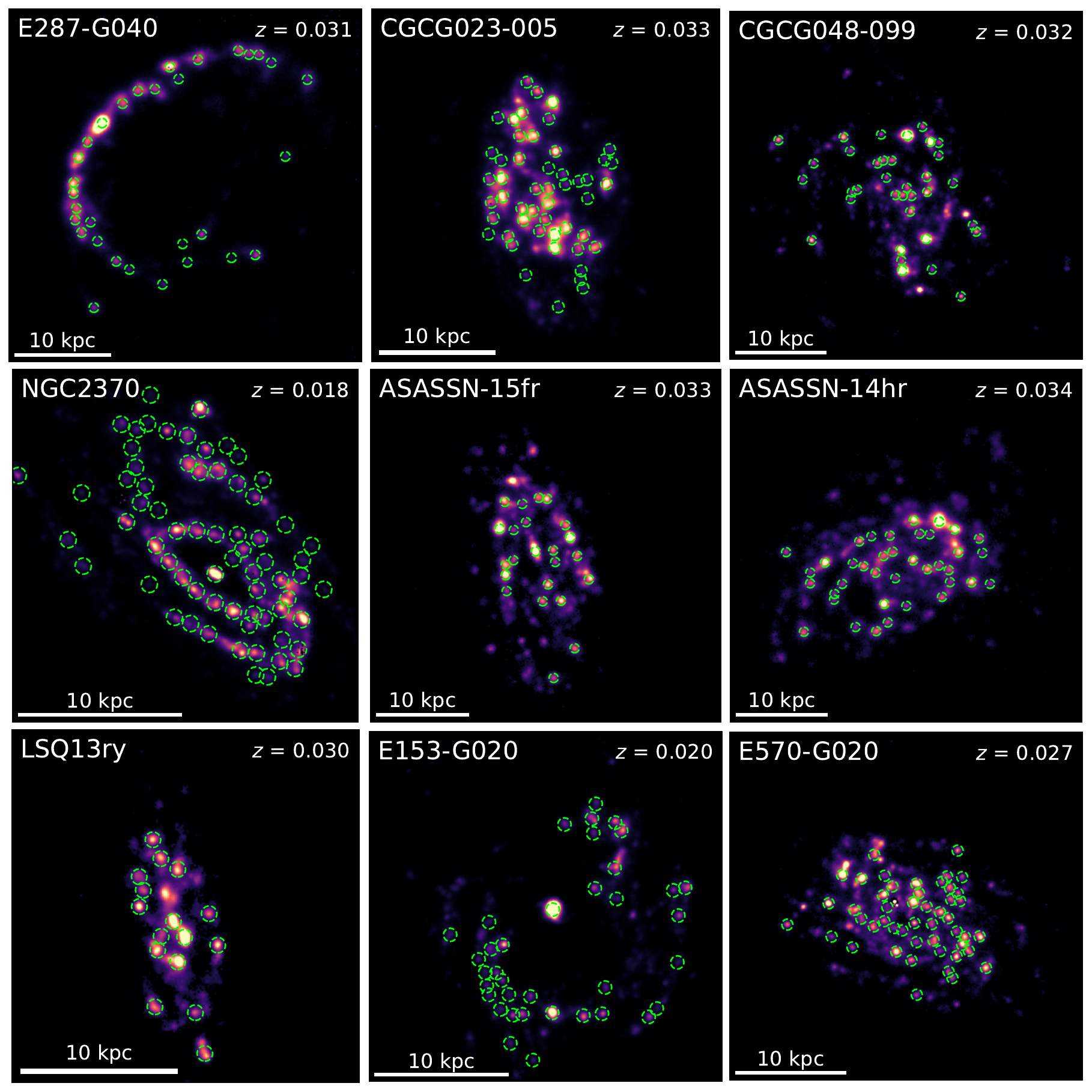}
    \caption{H$\alpha$ emission-line intensity maps from MUSE IFU spectroscopy of the galaxies used for our local calibration of the $L-\sigma$ relation. The apertures used to extract the \ion{H}{2} regions are indicated with green dashed circles. We extract the spectra of 347 total \ion{H}{2} regions from these nine galaxies.}
    \label{fig:muse_galaxies}
\end{figure*}

Using these criteria, we select nine galaxies from the AMUSING survey for our analysis (see Table \ref{tab:muse_galaxies}). The typical physical spatial resolution for the MUSE observations of the AMUSING galaxies ($\sim$400~pc) is comparable to that of the OSIRIS observations of Sp1149 ($\sim$~300pc). We download the MUSE IFU spectroscopy of each galaxy in the optical spectral range (4600--9350~\AA) from the ESO Archive\footnote{http://archive.eso.org/scienceportal/home}. For each galaxy, we create an H$\alpha$ emission-line intensity map and use the {\tt HIIdentify} software \citep{Easeman_2023} to identify \ion{H}{2} regions. We extract spectra of a total of 347  \ion{H}{2} regions using circular apertures with $r=500$~pc, equivalent to the apertures employed to extract the spectra of the \ion{H}{2} regions in Sp1149 from the OSIRIS data. {\tt pPXF} is used to model and remove the underlying stellar continuum  (see Section~\ref{subsec:osiris_obs}). The H$\alpha$ intensity maps of the MUSE galaxies are shown in Figure~\ref{fig:muse_galaxies}.

\begin{table}[]
\ra{1.3}
\centering
\caption{MUSE Model Priors$^a$}
\begin{tabular}{@{}lll@{}}

Parameter&Prior&Bounds\\

\midrule

$\log_{10}({\rm L_{H\alpha}})$ & Uniform & [36, 44] erg s$^{-1}$\\

$\sigma$ & Uniform &[0, 250] km s$^{-1}$\\

$\lambda_{\rm o,H\alpha}$ & Uniform & [$-10$, $+10$] $\lambda_{\rm H\alpha}(1+z)$~\AA\\

$A_V$ & Uniform &[0, 10] Magnitudes\\

$C_{\rm H\alpha}$, $C_{\rm H\alpha}$ & Uniform &[-1, 1] $10^{-16}$ erg s$^{-1}$ cm$^{-2}$ \AA$^{-1}$ \\

\bottomrule
\end{tabular}
{\footnotesize $^a$Prior constraints placed on each of the parameters used in the model for the MUSE data of local \ion{H}{2} regions. We use uniform priors with large parameter spaces. $\lambda_{\rm H\alpha}$ is the rest wavelength of H$\alpha$ ($\lambda_{\rm H\alpha}=6562.79$~\AA.)}
\label{tab:muse_priors}
\end{table}

The MUSE spectral wavelength coverage includes the H$\alpha$ and H$\beta$ emission lines of the \ion{H}{2} regions in the selected low-redshift galaxies. Using the inference method in Sections~\ref{subsec:Av_measurements} and \ref{subsec:osiris_measurements}, we measure the H$\alpha$ luminosity and intrinsic velocity dispersion of each \ion{H}{2} region in the local sample. Our model includes six parameters: the logarithm of the intrinsic luminosity of H$\alpha$ ($\log_{10}({\rm L_{H\alpha}})$), the intrinsic velocity dispersion ($\sigma$), the observed central wavelength of the H$\alpha$ emission line ($\lambda_{\rm o,H\alpha}$), the extinction due to dust ($A_V$), and the flux density of the stellar continuum near the H$\beta$ and H$\alpha$ lines ($C_{\rm H\beta}$ and $C_{\rm H\alpha}$, respectively). We use a uniform prior spanning a broad range of possible values for each of these parameters (see Table~\ref{tab:muse_priors}). 

From a set of model parameters, we compute models for the spectra spanning the H$\beta$ and H$\alpha$ lines for each \ion{H}{2} region and compare them to the observed MUSE spectrum to infer the best-fit parameter values. To convert from the intrinsic H$\alpha$ luminosity to the observed flux of H$\beta$ and H$\alpha$, we assume Case B recombination and use the \cite{Cardelli_1989} extinction law with $R_V=3.10$. We repeat our analysis with $R_V=2.51$ and find that the choice of $R_V$ does not significantly impact our results (see Section \ref{subsec:systematics}). We compute the total dispersion of the emission-line profiles using Eq.~\ref{equation:dispersion}, with $T_e=8000$~K as the typical electron temperature of the \ion{H}{2} regions. The central wavelength of the H$\alpha$ emission line is given by the parameter $\lambda_{\rm obs,H\alpha}$, and the central wavelength of the H$\beta$ profile is  $\lambda_{\rm obs,H\beta} = \lambda_{\rm obs,H\alpha} \times 4861/6563$.

Using the flux, width, and central wavelength of each emission line, we compute the model flux density as a function of wavelength (Eq.~\ref{equation:line_model}), and compute the likelihood  (Eq.~\ref{equation:likelihood}) for each \ion{H}{2} region. As for our analysis of the Keck spectra, we sample from the parameter space using an MCMC sampling method with the {\tt pymc} software. We obtain 10,000 samples after a burn-in of 5000 steps. 

The best-fit values and uncertainties for the intrinsic H$\beta$ luminosity and velocity dispersion of each \ion{H}{2} region in the local sample are used to calibrate the $z\approx 0$ $L-\sigma$ relation. We apply a hierarchical Bayesian linear regression algorithm using the {\tt linmix} software software \citep{Kelly_2007} to infer the slope and intercept of the local $L-\sigma$ relation,
\begin{equation}\label{equation:Lsigma}
    \log(L_{\rm{H}\alpha}) = (0.85\pm0.11) \log(\sigma) + (38.46\pm0.13)\, ,
\end{equation}
\noindent where $L_{H\alpha}$ is in units of erg s$^{-1}$ and $\sigma$ is in units of km s$^{-1}$. The root-mean-square intrinsic scatter is 0.34 dex. The H$\alpha$ luminosities and velocity dispersions of the local \ion{H}{2} regions, along with our fit for the $L-\sigma$ relation, are shown in Figure~\ref{fig:sp1149_Lsigma}.

We note that two of the galaxies in our low-$z$ sample, 2MASSX\,J01504127 and E287-G040, have poorer physical spatial resolution (790~pc and 920~pc, respectively) compared to the other galaxies in the sample and the typical resolution of the OSIRIS observations of the \ion{H}{2} regions in Sp1149. To test whether the poorer spatial resolution impacts our calibration, we repeat the linear fit to the velocity dispersions and H$\alpha$ luminosities of the low-redshift \ion{H}{2} regions with the regions from 2MASSX\,J01504127 and E287-G040 removed from the sample. We find that removing the galaxies with the poorest spatial resolution from the sample does not affect the inferred slope and intercept of the low-redshift $L-\sigma$ calibration. 

\section{Results and Discussion}\label{sec:results}
\subsection{The $L-\sigma$ Relation in Sp1149 at $z=1.49$}\label{subsec:Lsigma_results}

\begin{figure*}[ht]
    \centering
    \includegraphics[width=.9\linewidth]{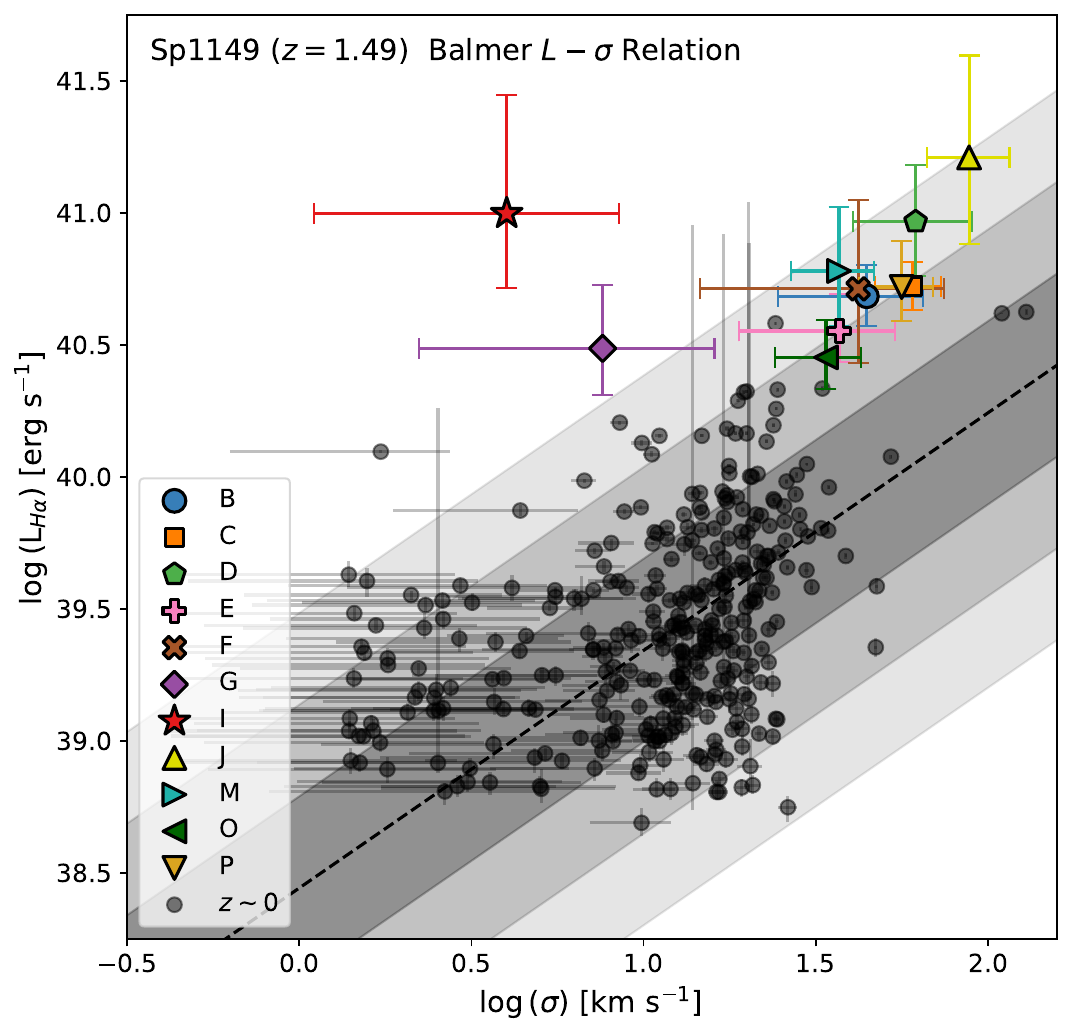}
    \caption{The velocity dispersions and intrinsic H$\alpha$ luminosities of 11 \ion{H}{2} regions at  $z=1.49$ in the magnified galaxy Sp1149 (colored points), compared to a low-redshift ($z\approx 0.03$) calibration of the $L-\sigma$ relation made using the same physical aperture sizes and similar spectral resolution (black points; gray shading represents the 1$\sigma$, 2$\sigma$, and 3$\sigma$ scatter of the local calibration).}
    \label{fig:sp1149_Lsigma}
\end{figure*}

We find that the \ion{H}{2} regions in Sp1149 have H$\alpha$ luminosities that are $\sim6$ times brighter than predicted by the locally calibrated $L-\sigma$ relation (see Figure~\ref{fig:sp1149_Lsigma}). Excluding the anomalously bright Region I (see Section~\ref{subsec:regionI}), we use the same linear regression method as described in Section~\ref{sec:calibration} to fit for the slope of the $L-\sigma$ relation among ten \ion{H}{2} regions in Sp1149, constraining the slope to be $0.86 \pm 5.24$. The \ion{H}{2} regions in Sp1149 are each 2$\sigma$--4$\sigma$ more luminous than the calibration of \ion{H}{2} regions at $z\approx 0$ predicts (see Table~\ref{tab:offsets}). 

\begin{table}[]
    \centering
    \ra{1.3}
    \setlength{\tabcolsep}{10pt}
    \caption{Sp1149 $L-\sigma$ Offsets$^a$}
    \begin{tabular}{ccc}
        
        \ion{H}{2} Region & $\Delta\log(L_{H\alpha}$ / erg s$^{-1}$) & Significance  \\
        \midrule
        B & 0.82 $\pm$ 0.36 & 2.3$\sigma$\\
        C & 0.75 $\pm$ 0.35 & 2.1$\sigma$\\
        D & 0.99 $\pm$ 0.40 & 2.5$\sigma$\\
        E & 0.76 $\pm$ 0.36 & 2.1$\sigma$\\
        F & 0.87 $\pm$ 0.44 & 2.0$\sigma$\\
        G & 1.28 $\pm$ 0.38 & 3.3$\sigma$\\
        I & 2.03 $\pm$ 0.44 & 4.6$\sigma$\\
        J & 1.10 $\pm$ 0.47 & 2.3$\sigma$\\
        M & 0.99 $\pm$ 0.40 & 2.5$\sigma$\\
        O & 0.69 $\pm$ 0.36 & 1.9$\sigma$\\
        P & 0.77 $\pm$ 0.36 & 2.1$\sigma$\\
        \bottomrule
    \end{tabular}
    {\footnotesize $^a$Differences between the measured H$\alpha$ luminosities of the \ion{H}{2} regions in Sp1149 and the predicted luminosities from the low-redshift $L-\sigma$ calibration at their velocity dispersions.}
    \label{tab:offsets}
\end{table}

We next fit a line to velocity dispersions and H$\alpha$ luminosities of the \ion{H}{2} regions in Sp1149, with the slope held fixed to the slope of the local $L-\sigma$ relation (0.85). The  $L-\sigma$ intercept for the \ion{H}{2} regions in Sp1149 is $39.27\pm0.06$, while the local $L-\sigma$ intercept is $38.46\pm0.13$ (see Eq.~\ref{equation:Lsigma}). Consequently, the intercept for Sp1149 would be offset from the local calibration by $0.81\pm0.16$~dex. The \ion{H}{2} regions in Sp1149 are therefore $6.4^{+2.9}_{-2.0}$ times more luminous than the  $L-\sigma$ relation at low redshift would predict, a 3$\sigma$ tension. 

There are two possible explanations for the observed tension. First, this result may suggest that \ion{H}{2} regions are more luminous for a given velocity dispersion in Sp1149 and thus a fundamental physical difference between giant \ion{H}{2} regions in this galaxy at $z=1.49$ and those at low redshift.  

Alternatively, the tension we find between between the magnified \ion{H}{2} regions in Sp1149 and those at $z\approx 0$ could arise from a systematic bias in the GLAFIC MACS\,J1149 cluster mass model. If the true magnification at the positions of the \ion{H}{2} regions were a factor of $\sim6$ times smaller than predicted by the {\tt GLAFIC} model, then our analysis would overestimate the H$\beta$ luminosities of the magnified \ion{H}{2} regions by the same factor. In this case, the \ion{H}{2} regions in Sp1149 would be consistent with the $L-\sigma$ relation at low redshift. In our companion paper, Williams et al. (2023, submitted), we evaluate ten available models of the MACS\,J1149 cluster which were created by different teams using varying modeling techniques and distinct combinations of input constraints. For all ten models, we find that the magnification-corrected H$\beta$ luminosities of the \ion{H}{2} regions in Sp1149 are $\sim5$--8 times brighter than expected by the local $L-\sigma$ relation. Therefore, in order for a bias in the magnification maps of the MACS\,J1149 cluster to explain the observed $L-\sigma$ tension, the same bias would have to apply to all ten of these models. Additionally, we find that the flux ratios of bright knots in Image 1.1 and Image 1.3 are consistent with the predicted magnification ratios at their positions (Williams et al. 2023, submitted), suggesting that the predicted magnifications in Image 1.1 are likely not overestimated. 

The lens modeling methods used to create the MACS\,J1149 cluster models have been tested using simulated images of clusters with the same resolution and depth of the HFF survey. \cite{Meneghetti_2017} found that most lens models can reliably predict the magnification within an accuracy of $\sim30\%$ for magnifications $\mu \approx 10$. We therefore expect that the observed tension is more likely to be caused by a true physical difference in the \ion{H}{2} regions in Sp1149 compared to those in our low-redshift sample rather than to a bias in the magnification maps of MACS\,J1149. 

\subsection{BPT Diagram}\label{subsec:BPT_results}

Using our measurements on the emission-line fluxes from the MOSFIRE spectroscopy of the \ion{H}{2} regions in Sp1149, we constrain the emission-line flux ratios of $F$([\ion{O}{3}])/$F$(H$\beta$) (denoted O3) and $F$([\ion{N}{2}])/$F$(H$\alpha$) (denoted N2). The position of a galaxy on the BPT diagram reflects its dominant source of ionizing radiation. Galaxies that host active galactic nuclei (AGNs) have high ratios of both O3 and N2, while galaxies that are dominated by low-ionization nuclear emission-line regions (LINERs) have high N2 but lower O3 compared to AGNs. SFGs and \ion{H}{2} regions at $z\approx 0$ fall on a well-defined locus on the BPT diagram \citep{Kewley_2013}, 

\begin{equation}
    {\rm O3} = \left(\frac{0.61 }{\rm{N2} + 0.08} \right) + 1.1\, ,
\end{equation}
\noindent
and SFGs at $z\approx 2$ follow a curve that is offset from the $z\approx 0$ locus by $\Delta {\rm N2} = 0.37$~dex \citep[e.g.,][]{Strom_2017}. Possible explanations for this offset include larger ionization parameters at higher redshifts \citep{Kewley_2013,Kashino_2017} and stellar $\alpha$-enhancement at higher redshifts producing harder ionizing spectra at fixed metallicity \citep{Steidel_2016,Strom_2018,Shapley_2019,Sanders_2020}.  The offset of high-redshift galaxies may also be caused by a sample selection effect; the galaxies observed at $z\approx 2$ are more likely to have higher stellar masses ($\log(M_*/M_\odot)\approx 10$--11) and therefore higher metallicities, driving them to higher N2 and O3 values on the BPT diagram \citep{Garg_2022}.

\begin{figure}[]
    \centering
    \includegraphics[width=\linewidth]{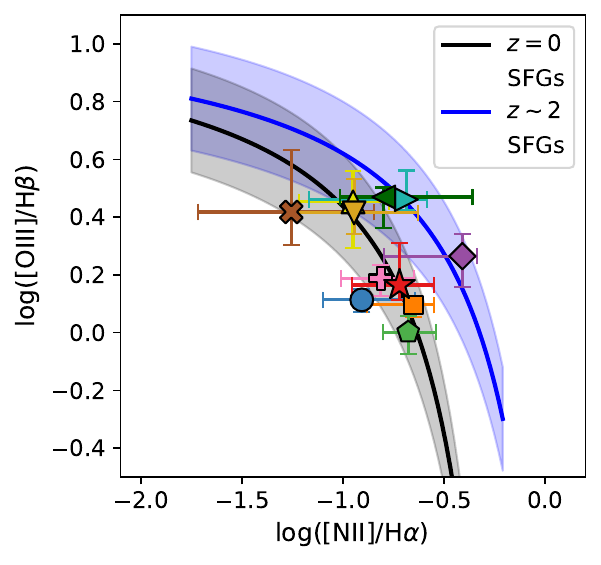}
    \caption{BPT diagram for the \ion{H}{2} regions in Sp1149. Plotting symbols are the same as Figure~\ref{fig:sp1149_Lsigma}. The fit for the $z=0$ SFG locus is from \cite{Kewley_2013} and the fit for the $z=2$ SFG locus is from \cite{Strom_2017}. We find that the majority of the \ion{H}{2} regions we measured in Sp1149 follow the $z=0$ SFG locus within the 1$\sigma$ uncertainties.}
    \label{fig:BPT}
\end{figure}

We find that the 11 \ion{H}{2} regions we measured in Sp1149 are consistent with the $z=0$ BPT locus within the 1$\sigma$ uncertainties, indicating that the regions we used in our analysis are dominated by ionizing radiation from hot young  stars (see Figure~\ref{fig:BPT}). Since the stellar mass of Sp1149 \citep[$\log(M_*/M_\odot) = 9.64^{+0.05}_{-0.01}$;][]{Wang_2017}  is relatively small compared to the masses of the sample of $z\approx 2$ galaxies that are offset from local galaxies on the BPT diagram, this result supports the idea that a sample selection effect may contribute to the $z\approx 2$ BPT offset. Alternatively, at $z=1.49$, Sp1149 may have a lower level of $\alpha$-enhancement compared to galaxies at $z\approx2$. We note that the presence of a bright night-sky line near the observer-frame wavelength of [\ion{N}{2}] (see Figure~\ref{fig:mosfire_profiles}) limits our ability to constrain precisely the flux of [\ion{N}{2}] in the \ion{H}{2} regions in Sp1149, leading to large uncertainties in the measured N2 ratios (see Figure~\ref{fig:BPT}).  

\subsection{An Anomalously Bright \ion{H}{2} Region}\label{subsec:regionI}

\ion{H}{2} Region I in Sp1149 is extremely luminous given its very small intrinsic velocity dispersion ($\log(L_{\rm H\alpha}$/(erg~s$^{-1}$)) = $41.00^{+0.45}_{-0.28}$; $\sigma = 4.00^{+4.49}_{-2.90}$ km s$^{-1}$). It is $2.03\pm0.44$ dex ($\sim5\sigma$) brighter than the prediction from the locally calibrated $L-\sigma$ relation at 4.00 km s$^{-1}$. Compared to the other \ion{H}{2} regions in Sp1149 with similar luminosities, the small velocity dispersion of Region I is a 3.5$\sigma$ outlier.   

In principle, the extremely high H$\alpha$ luminosity might be explained by a magnification anomaly, such as millilensing by a subhalo, at the position of \ion{H}{2} Region I. To test this possibility, we obtained follow-up \textit{H}-band spectroscopy of Region I in both Image 1.1 and Image 1.2 of Sp1149 using MOSFIRE on November 15, 2022 (see Section~\ref{subsec:MOSFIRE}). We measured the flux of the H$\alpha$ emission line in both images of Region I, and found a flux ratio between Image 1.1 and Image 1.2 of $1.40 \pm 0.26$. The ratio of the magnifications predicted by the GLAFIC model at the image positions is $1.70 \pm 0.30$, so the observed H$\alpha$ flux ratio between the images is consistent with the predicted magnification ratio within the 1$\sigma$ uncertainties. The observed H$\alpha$ flux of Region I in Image 1.1 is also consistent between the 2020 and 2022 MOSFIRE observations (see Table~\ref{tab:regionI}). These results suggest that the bright H$\alpha$ luminosity observed in Region I is not caused by millilensing.

Another possible explanation is that the source itself is responsible for the unusually high H$\alpha$ luminosity we measure in Region I. Even though it is not coincident with the galaxy nucleus, we consider the possibility that it could come from an AGN, LINER, or some other source with narrow emission lines. If the source were an AGN or a LINER, we would expect its emission-line ratios to fall above the star-forming locus of the BPT diagram. Region I's emission-line ratios are consistent, however, with the locus of star-forming galaxies (SFGs) in the BPT diagram (see Figure~\ref{fig:BPT}), so it is unlikely that the source is an AGN or a LINER. 

We note that the Balmer decrement for Region I are uncertain yet also consistent with the possiblity of high extinction, $A_V=1.43^{+1.33}_{-0.82}$~mag. If the extinction is overestimated, then the intrinsic luminosity would also be overestimated, which could in principle contribute to the observed tension. To test this possibility, we compute the intrinsic H$\alpha$ luminosity of Region I with the extinction fixed to the mean extinction of the \ion{H}{2} regions in Sp1149, $A_V=0.79\pm0.53$~mag. We find $\log(L_{\rm H \alpha}$/(erg~s$^{-1}$)) = $40.78\pm0.20$, which is 0.22~dex fainter than our original measurement yet is still $\sim 65$ times brighter than the local $L-\sigma$ relation would predict given its velocity dispersion. 

\begin{figure*}[ht]
    \centering
    \includegraphics[]{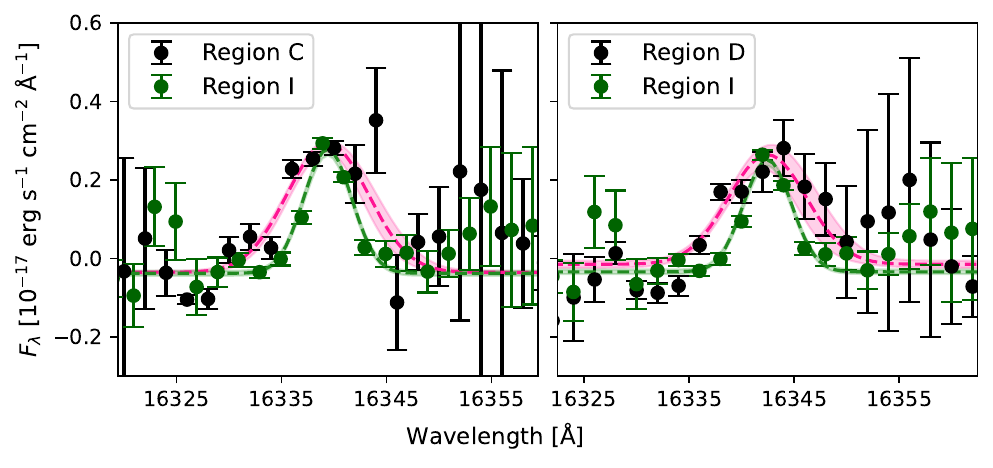}
    \caption{The narrow OSIRIS H$\alpha$ profile of Region I, shown on top of the OSIRIS H$\alpha$ profiles of \ion{H}{2} regions with similar intrinsic Balmer luminosities (Regions C and D). The spectrum of Region I has been renormalized and recentered in each panel to match the profile of the other \ion{H}{2} region. We find that the H$\alpha$ emission profile of Region I is narrower than that of Regions C and D by $\sim 3\sigma$.   }
    \label{fig:regI_width}
\end{figure*}

\begin{table}[]
    \centering
    \ra{1.3}
    \caption{Region I Flux Measurements$^a$}
    \begin{tabular}{ccc}
        \multicolumn{3}{c}{\textbf{Region I Flux Measurements}}\\
        Observation & $F_{\rm H\alpha}$ ($10^{-17}$ erg s$^{-1}$ cm$^{-2}$)  & $\mu$  \\
        \midrule
        Img 1.1 (2020)& $6.43\pm0.66$ &$12.3\pm0.7$\\
        Img 1.1 (2022) & $6.63\pm0.82$ & $12.3\pm0.7$ \\
        Img 1.2 (2022) & $4.72\pm0.64$ & $7.2\pm1.2$ \\
        Ratio (1.1/1.2) & $1.40\pm0.26$ & $1.70 \pm 0.30$\\
        \bottomrule
    \end{tabular}
    {\footnotesize $^a$MOSFIRE H$\alpha$ flux measurements of \ion{H}{2} Region I in Images 1.1 and 1.2 of Sp1149. The H$\alpha$ fluxes in Image 1.1 are consistent between observations, and the flux ratio between Image 1.1 and Image 1.2 is consistent with the predicted magnification ratio at the image positions. }
    \label{tab:regionI}
\end{table}

\subsection{Consistency Checks}\label{subsec:systematics}
To test for systematic errors, we repeat our analysis using additional dust extinction curves and aperture sizes. In our initial analysis, we used an $R_V=3.10$ extinction curve for the dust extinction correction in our low-redshift sample and the Sp1149 \ion{H}{2} regions. We repeat the low-redshift calibration and the Sp1149 $L$ and $\sigma$ measurements using $R_V=2.51$, the typical value for SFGs at $z\approx 2$ from the MOSDEF survey \citep{Reddy_2015}. We find that the inferred luminosities of the \ion{H}{2} regions in Sp1149 are at most 20\% lower with $R_V=2.51$ compared to $R_V=3.10$. The inferred offset from the intercept of the low-redshift $L-\sigma$ relation is $0.85 \pm 0.16$~dex with $R_V=2.51$, consistent with the offset measured with $R_V=3.10$ ($0.81\pm0.16$).

We used 500~pc circular apertures in our original analysis to extract the spectra of the \ion{H}{2} regions in Sp1149 and the low-redshift sample. We repeat our analysis using 400~pc and 600~pc apertures to extract the spectra of both samples. We find that the inferred offset of the intercept of the $L-\sigma$ relation is $0.89 \pm 0.16$ for the 400~pc apertures, and $0.92\pm0.16$ for the 600~pc apertures, consistent within the 1$\sigma$ uncertainties with the offset measured using the 500~pc apertures ($0.81\pm0.16$).

\section{Conclusions}\label{sec:conclusion}
Using a combination of Keck OSIRIS AO together with Keck MOSFIRE and LBT LUCI natural-seeing spectroscopy, we measured the Balmer luminosities and velocity dispersions of 11 giant \ion{H}{2} regions in the largest image of the strongly lensed spiral galaxy Sp1149 at $z=1.49$. The OSIRIS AO spectroscopy provided a typical physical resolution of 300~pc in the source plane. We derived a low-redshift calibration of the Balmer $L-\sigma$ relation using measurements from MUSE spectroscopy of 347 \ion{H}{2} regions in nearby ($z\approx 0$) spiral galaxies, with similar aperture sizes and spectral resolution as we used in the OSIRIS analysis. 

We found that the \ion{H}{2} regions in the lensed galaxy at $z=1.49$ have H$\alpha$ luminosities that are a factor of $6.4^{+2.9}_{-2.0}$ brighter than the local $L-\sigma$ relation predicts for a given velocity dispersion, assuming a fixed slope. The difference could arise from a physical difference between the \ion{H}{2} regions in Sp1149 and those in the $z\approx 0.03$ sample. Alternatively, a systematic bias in the magnification predictions from the lens models of the MACS\,J1149 cluster might explain the observed tension. We show in our companion paper that  the predicted magnification ratios between the same knots seen in different images of Sp1149 are consistent with the observed flux ratios, so it is unlikely that the lens models are systematically biased. However, observations of \ion{H}{2} regions that are magnified by additional clusters with different lensing configurations would be needed to exclude this possibility. A larger sample of resolved \ion{H}{2} regions at $z\gtrsim1$ would also be necessary to determine whether their \ion{H}{2} regions are also substantially brighter than the low-redshift $L-\sigma$ relation would predict to determine if Sp1149 may be typical.

Using emission-line flux ratios that we measured from the MOSFIRE spectra, we found that the \ion{H}{2} regions in Sp1149 are consistent with the low-redshift star-forming locus on the BPT diagram, which indicates that the regions used in our analysis are photoionized by hot young stars. There is a well-known high-redshift offset in the star-forming locus of the BPT diagram, according to which SFGs at $z\approx 2$ having higher N2 compared to those at $z=0$. Interestingly, the \ion{H}{2} regions in Sp1149 ($z=1.49$) agree with the $z=0$ locus in the BPT diagram. Given that the stellar mass of Sp1149 is relatively small in comparison with that of the existing sample, this result may suggest that the observed high-redshift offset in the BPT diagram could be connected to a sample selection effect; galaxies used in the $z\approx 2$ samples typically have large stellar masses ($\log(M_*/M_\odot)\approx 10$--11), leading to higher metallicities and therefore higher N2 values \citep{Garg_2022}. We note, however, that the presence of a strong night-sky line at the observer-frame wavelength of [\ion{N}{2}] leads to large uncertainties in N2 for the \ion{H}{2} regions in Sp1149.  

One \ion{H}{2} region in Sp1149, Region I, is very luminous ($\log(L_{\rm H\alpha}$/(erg~s$^{-1}$)) = $41.00^{+0.45}_{-0.28}$) given its small intrinsic velocity dispersion ($\sigma = 4.00^{+4.49}_{-2.90}$ km s$^{-1}$), approximately 110 times brighter than the $L-\sigma$ relation would predict. The H$\alpha$ flux ratio of the counterimages of Region I is consistent with the predicted magnification ratios, suggesting that the high inferred Balmer luminosity is not caused by millilensing by a subhalo. Region I is consistent with stellar photoionization on the BPT diagram, so it is unlikely that the source is an AGN or a LINER. The nature of this luminous source with very narrow emission lines warrants further investigation. 

\section{Acknowledgments}\label{sec:acknowledgments}
Some of the data presented herein were obtained at the W. M. Keck
Observatory, which is operated as a scientific partnership among the
California Institute of Technology, the University of California, and
the National Aeronautics and Space Administration
(NASA); the observatory was made possible by the generous financial
support of the W. M. Keck Foundation.
We thank Josh Walawender and Sherry Yeh for their support with the MOSFIRE observations and data reduction, and Tiantian Yuan for providing the reduced OSIRIS datacube. We obtained Keck data from telescope time allocated to A.V.F. through the University of California, and  to P.K. through NASA [Keck PI Data Award \#1644110 (75\/2020A\_N110), administered by the NASA Exoplanet Science Institute through the agency's scientific partnership with the California Institute of Technology and the University of California].  We recognize the Maunakea summit as a sacred site within the indigenous Hawaiian community and we are grateful for the opportunity to conduct observations there. 

P.K. acknowledges funding from NSF grant AST-1908823. A.V.F. was supported by the Christopher R. Redlich Fund and many individual donors.

\bibliography{main}{}
\bibliographystyle{aasjournal}

\clearpage

\appendix
\begin{figure*}[ht]
    \centering
    \includegraphics[width=.87\linewidth]{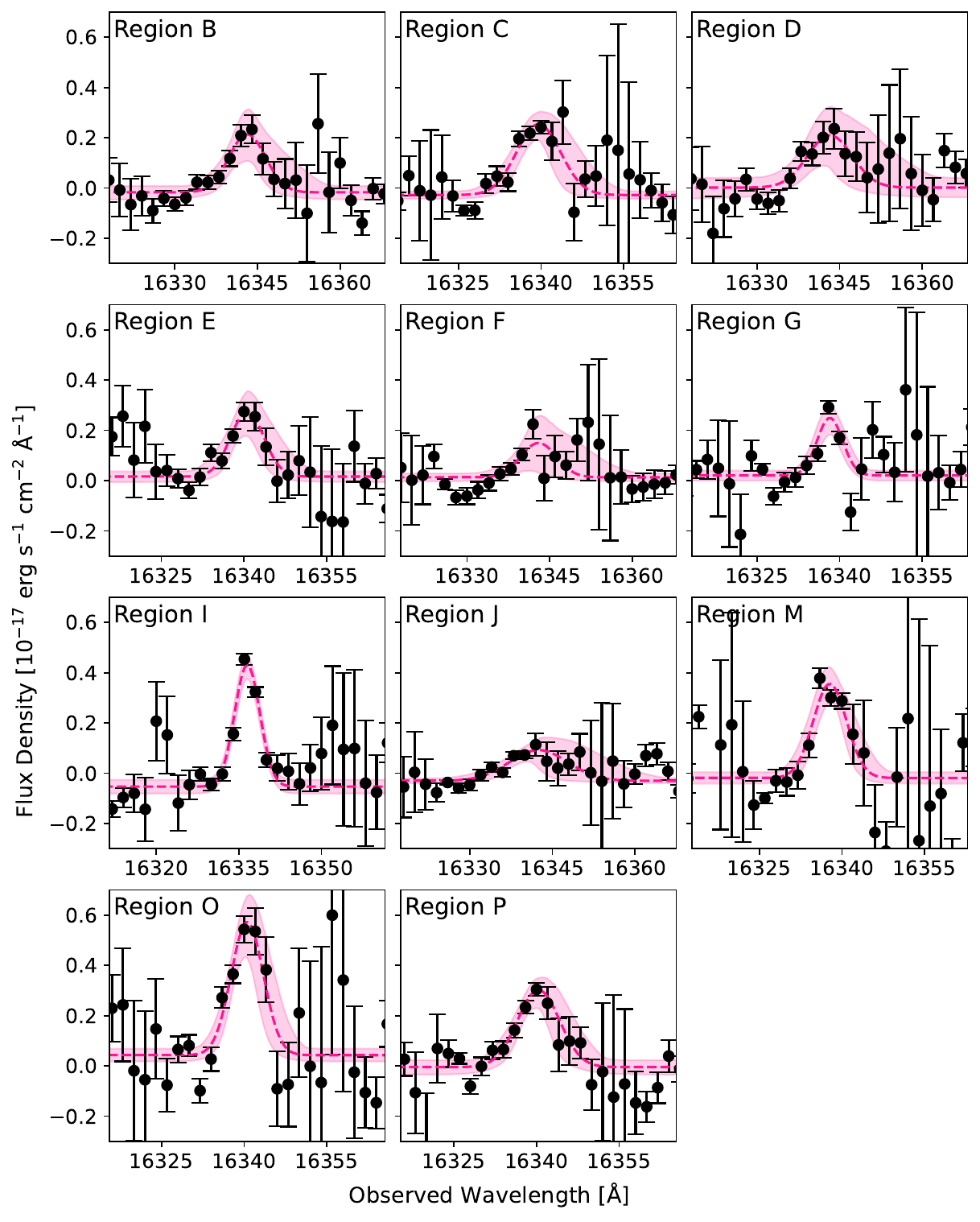}
    \caption{Best-fitting Gaussian profiles (pink dashed lines) on top of the OSIRIS spectrum of the H$\alpha$ line (black points) of each \ion{H}{2} region.}
    \label{fig:osiris_profiles}
\end{figure*}

\begin{figure*}[ht]
    \centering
    \includegraphics[width=.925\linewidth]{ 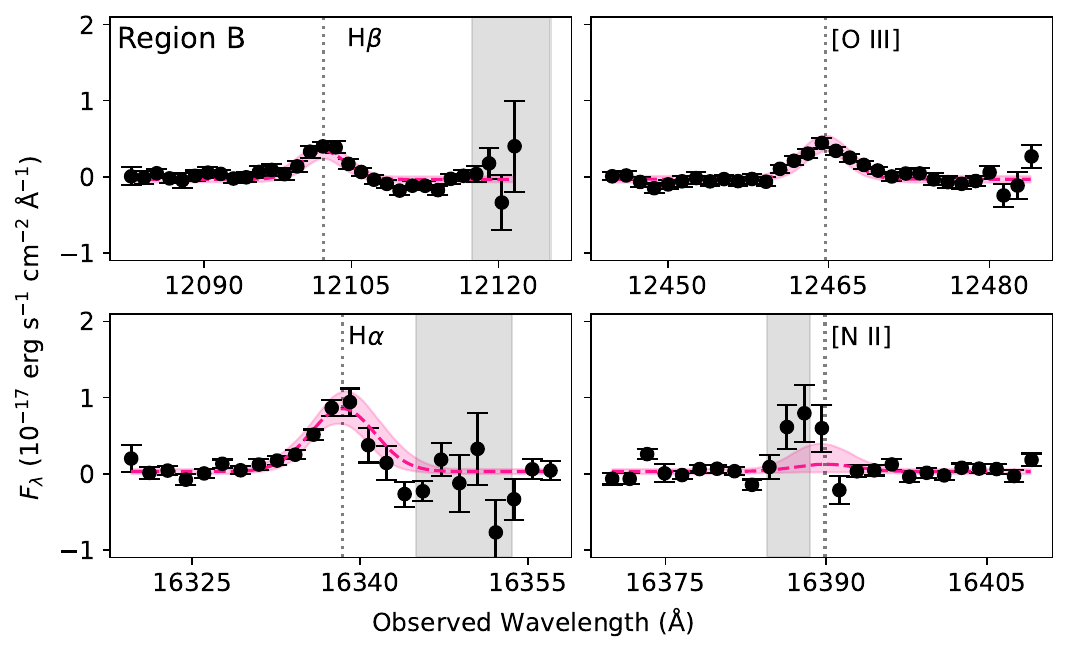}
    \caption{MOSFIRE H$\beta$, [\ion{O}{3}], H$\alpha$, and [\ion{N}{2}] emission-line profiles for each of the \ion{H}{2} regions in Sp1149. The black points show the observed MOSFIRE spectrum and the pink lines show the best-fit emission-line profiles. The gray shaded regions show the positions of prominent night-sky emission lines.}
    \label{fig:mosfire_profiles}
\end{figure*}

\begin{figure*}
    \centering
    \includegraphics[width=.95\linewidth]{ 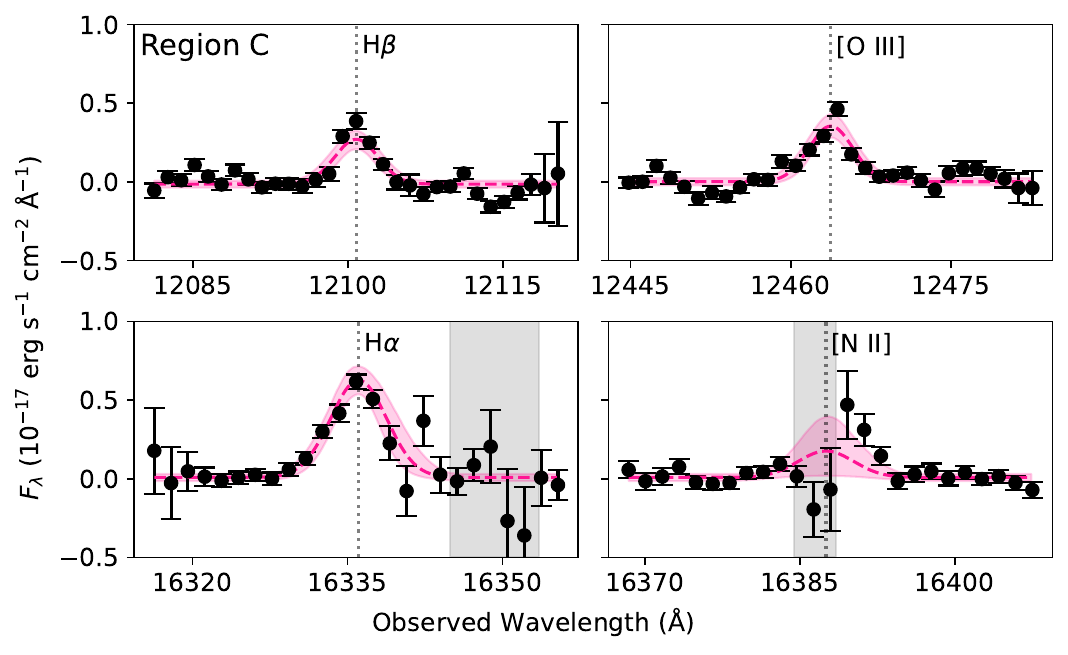}
    \textbf{Figure 10.} Continued
\end{figure*}

\begin{figure*}
    \centering
    \includegraphics[width=.95\linewidth]{ 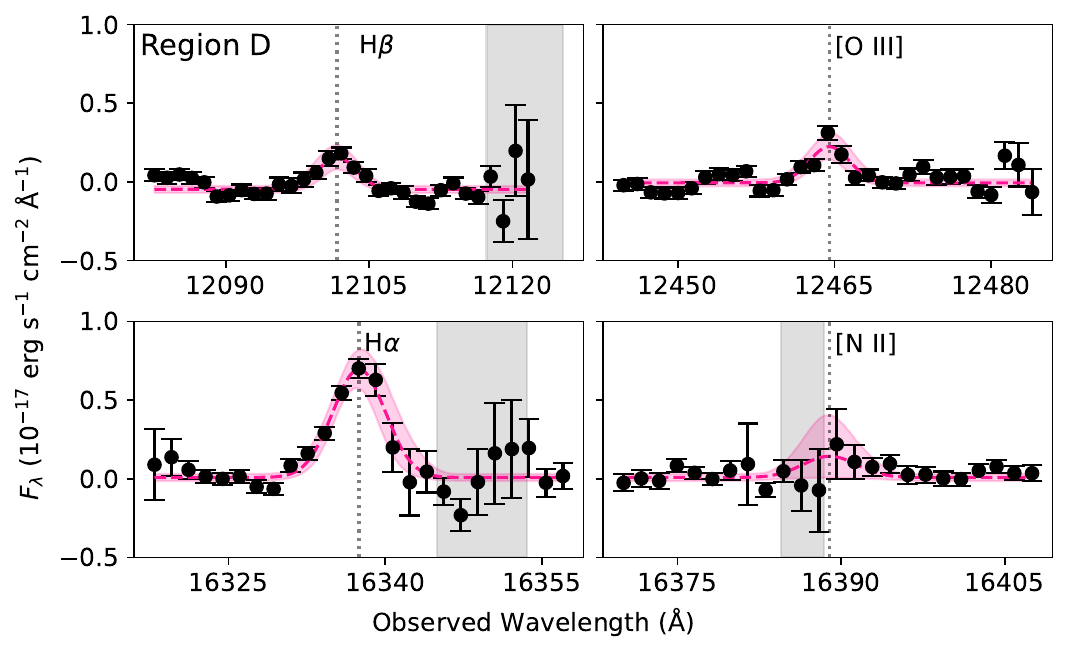}
    \textbf{Figure 10.} Continued
\end{figure*}

\begin{figure*}
    \centering
    \includegraphics[width=.95\linewidth]{ 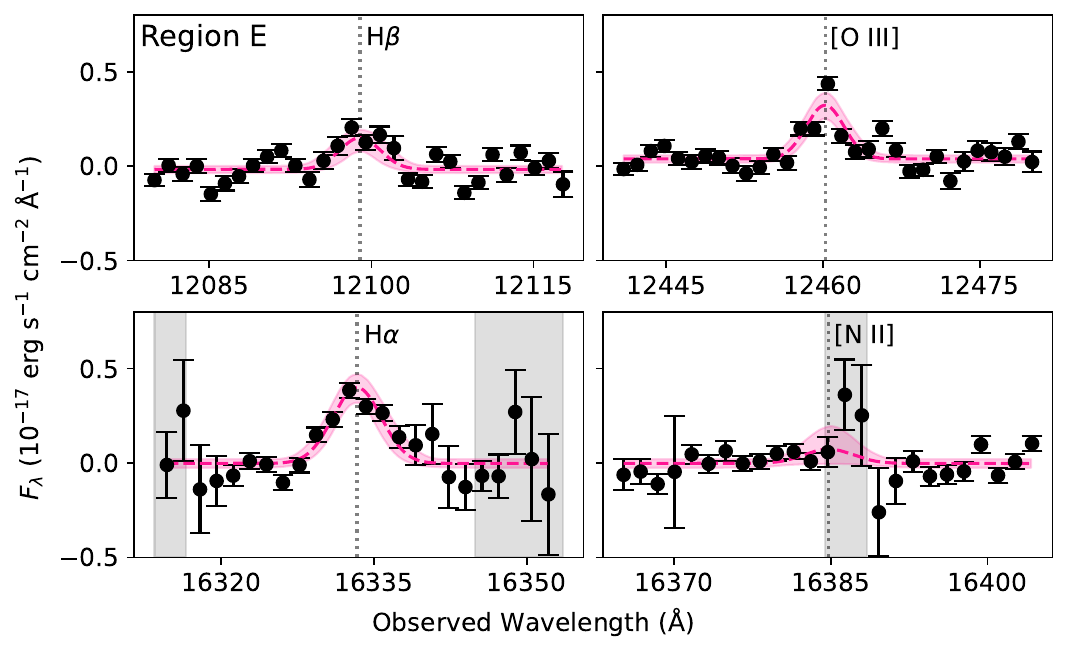}
    \textbf{Figure 10.} Continued
\end{figure*}

\begin{figure*}
    \centering
    \includegraphics[width=.95\linewidth]{ 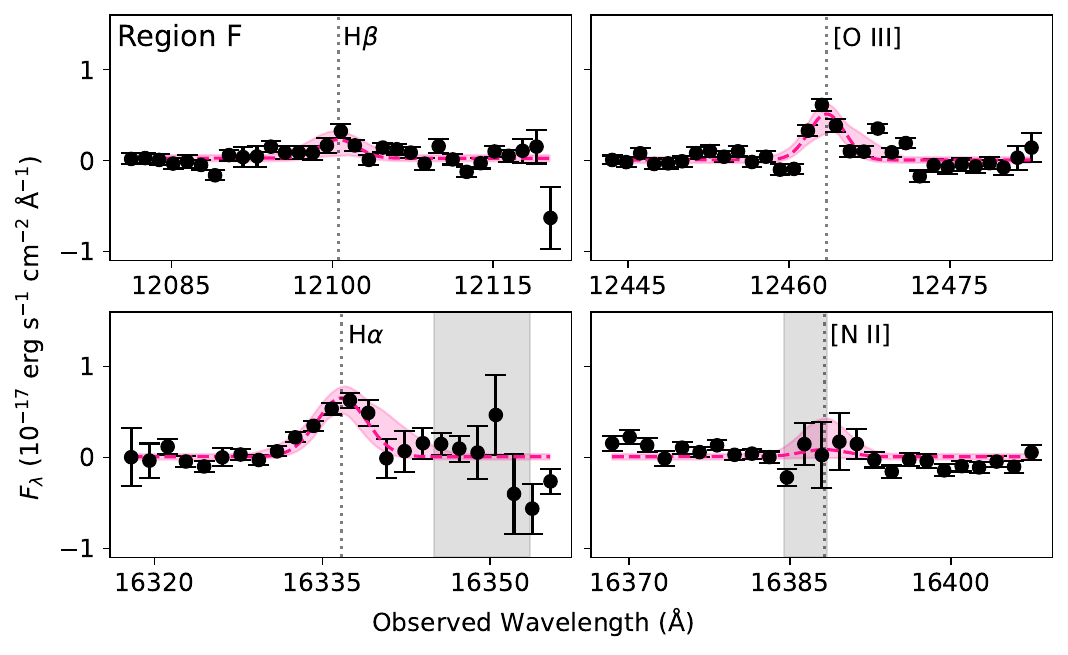}
    \textbf{Figure 10.} Continued
\end{figure*}

\begin{figure*}
    \centering
    \includegraphics[width=.95\linewidth]{ 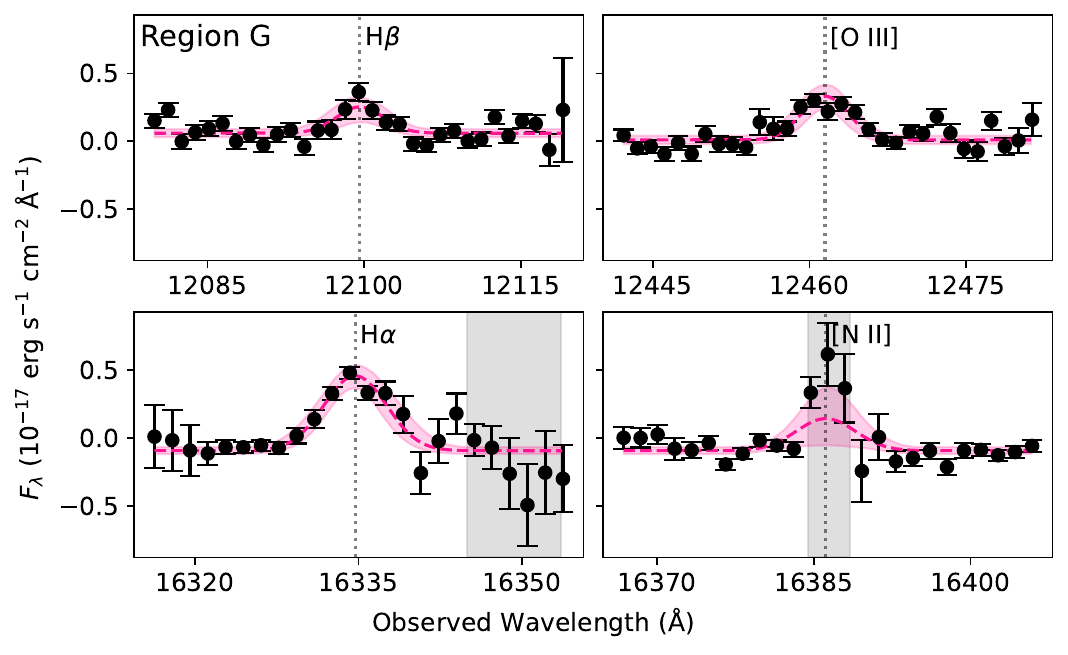}
    \textbf{Figure 10.} Continued
\end{figure*}

\begin{figure*}
    \centering
    \includegraphics[width=.95\linewidth]{ 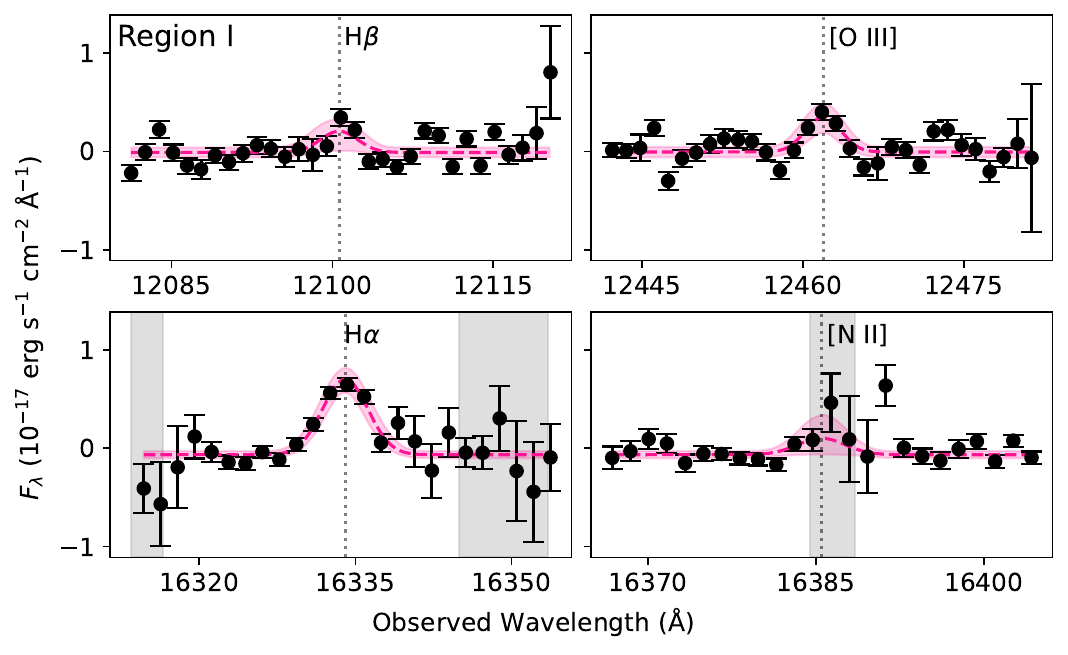}
    \textbf{Figure 10.} Continued
\end{figure*}

\begin{figure*}
    \centering
    \includegraphics[width=.95\linewidth]{ 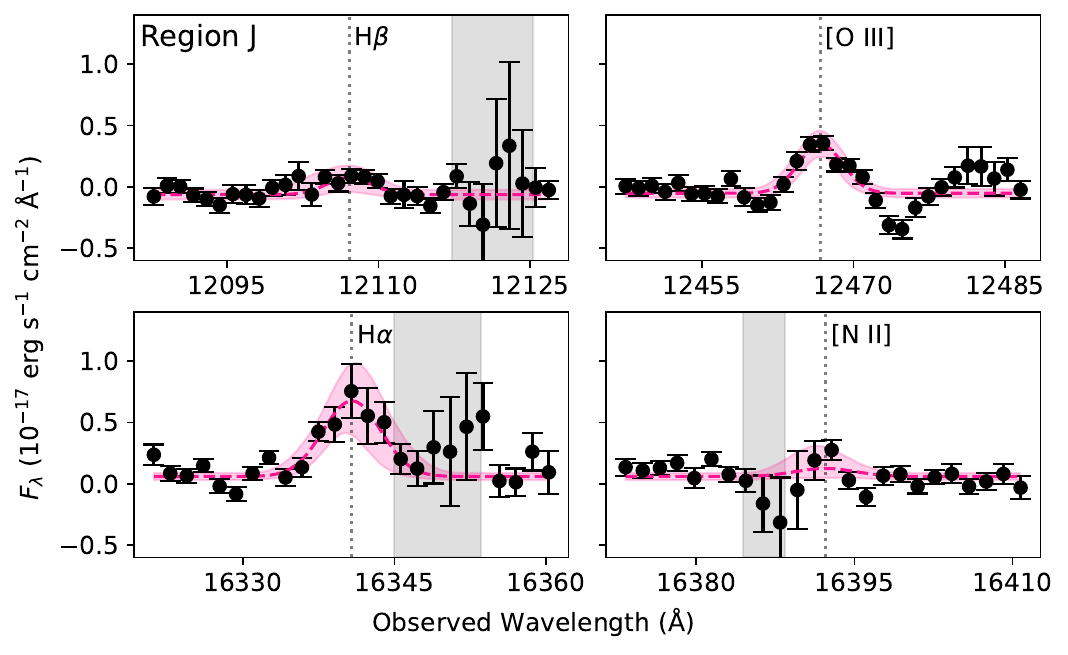}
    \textbf{Figure 10.} Continued
\end{figure*}

\begin{figure*}
    \centering
    \includegraphics[width=.95\linewidth]{ 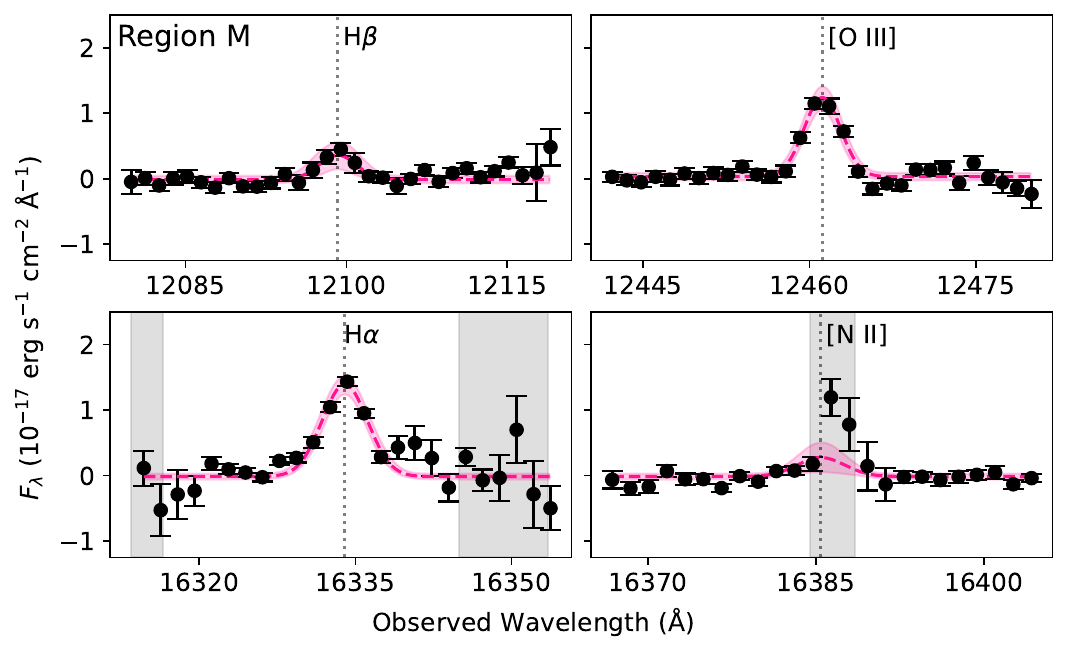}
    \textbf{Figure 10.} Continued
\end{figure*}

\begin{figure*}
    \centering
    \includegraphics[width=.95\linewidth]{ 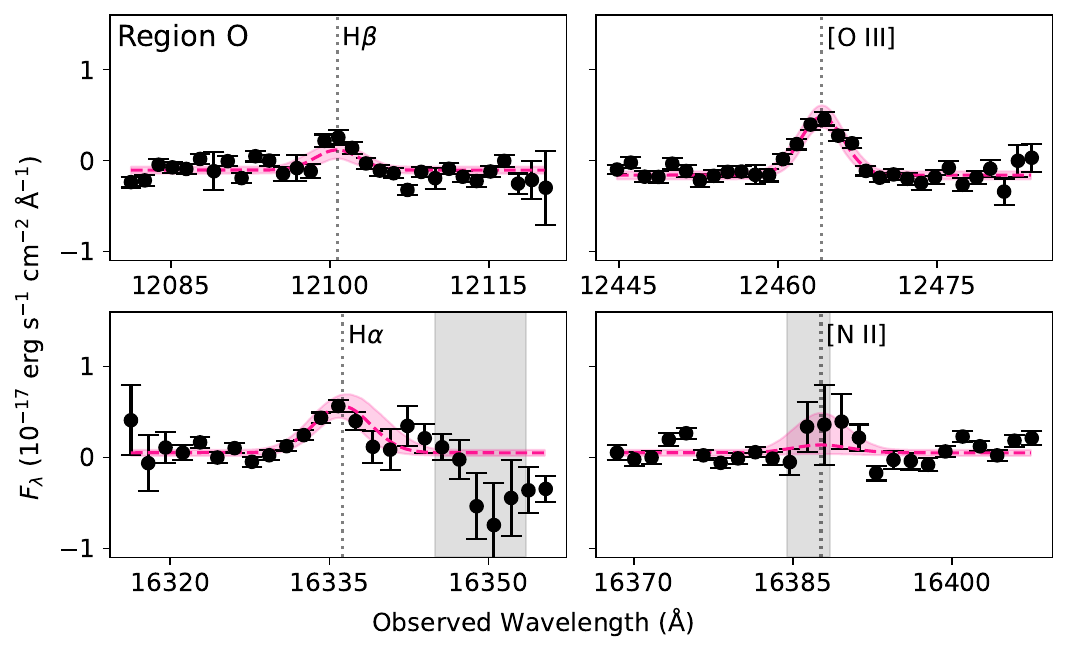}
    \textbf{Figure 10.} Continued
\end{figure*}

\begin{figure*}
    \centering
    \includegraphics[width=.95\linewidth]{ 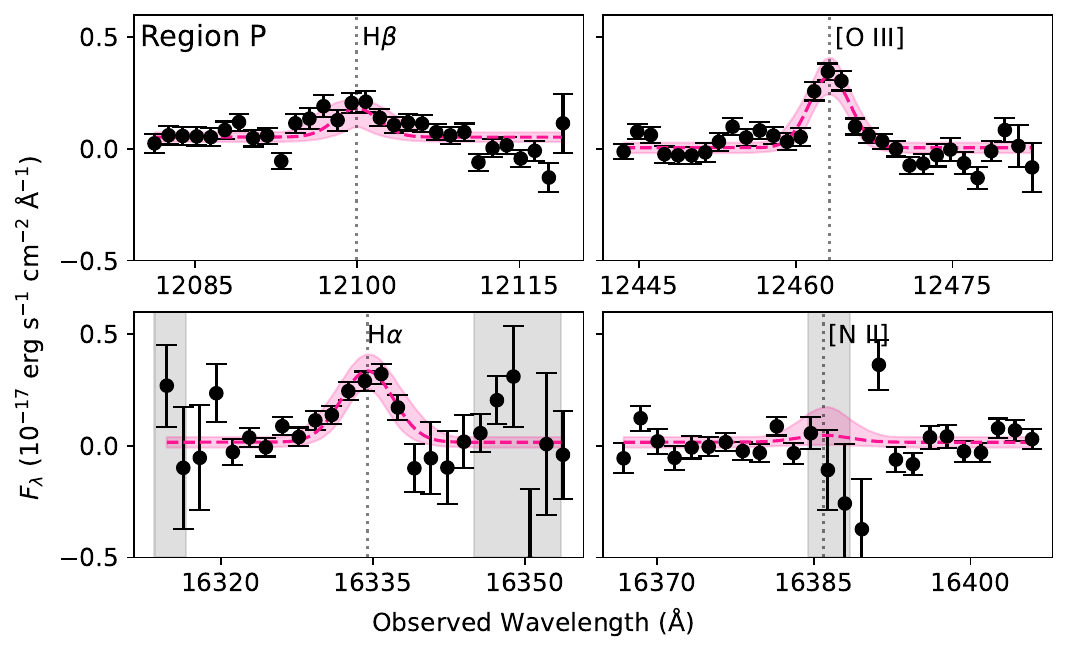}
    \textbf{Figure 10.} Continued
\end{figure*}

\end{document}